\documentclass[12pt,preprint]{aastex}

\usepackage{graphicx}
\usepackage{amssymb}
\usepackage{amsmath}
\usepackage{epstopdf}
\usepackage{natbib}
\usepackage{color}
\usepackage{hyperref}
\hypersetup{colorlinks=true, linkcolor=blue, citecolor=blue,urlcolor=blue}
\usepackage[all]{hypcap}
\usepackage[flushleft]{threeparttable}

\def\eqlbl#1{\label{eq:#1}}
\def\eqref#1{(\ref{eq:#1})}

\shorttitle{Stagnant lid planet habitability}
\shortauthors{Foley}

\begin{document}


\title{Habitability of Earth-like stagnant lid planets: Climate evolution and recovery from snowball states}


\author{Bradford J. Foley}
\affil{Department of Geosciences, Pennsylvania State University, University Park, PA, 16802}
\email{bjf5382@psu.edu}



\begin{abstract}
Coupled models of mantle thermal evolution, volcanism, outgassing, weathering, and climate evolution for Earth-like (in terms of size and composition) stagnant lid planets are used to assess their prospects for habitability. The results indicate that planetary CO$_2$ budgets ranging from $\approx 3$ orders of magnitude lower than Earth's to $\approx 1$ order of magnitude larger, and radiogenic heating budgets as large or larger than Earth's, allow for habitable climates lasting 1-5 Gyrs. The ability of stagnant lid planets to recover from potential snowball states is also explored; recovery is found to depend on whether atmosphere-ocean chemical exchange is possible. For a ``hard" snowball with no exchange, recovery is unlikely, as most CO$_2$ outgassing takes place via metamorphic decarbonation of the crust, which occurs below the ice layer. However, for a ``soft" snowball where there is exchange between atmosphere and ocean, planets can readily recover. For both hard and soft snowball states, there is a minimum CO$_2$ budget needed for recovery; below this limit any snowball state would be permanent. Thus there is the possibility for hysteresis in stagnant lid planet climate evolution, where planets with low CO$_2$ budgets that start off in a snowball climate will be permanently stuck in this state, while otherwise identical planets that start with a temperate climate will be capable of maintaining this climate for 1 Gyrs or more. Finally, the model results have important implications for future exoplanet missions, as they can guide observations to planets most likely to possess habitable climates.   
\end{abstract}



\section{Introduction}
\label{sec:intro}

Determining the factors that allow long-lived, habitable climates to develop on rocky planets is a critical goal for astrobiology, especially in light of the large number of exoplanets that have been discovered in the last 20 years \citep[e.g.][]{Wright2011,Batalha2014,Burke2014}. In particular, such knowledge can be used to guide future observing resources to those planets most likely to harbor life. In order to host surface life which can be remotely observed, it is thought that a planet must lie within the habitable zone, such that liquid water can exist on the planet's surface \citep[e.g.][]{kasting1993}. However, even within the habitable zone, stabilization of climate via the carbonate-silicate cycle is necessary to prevent changes in solar luminosity, or variations in CO$_2$ outgassing rate, from driving climate towards uninhabitably hot or cold states \citep[e.g.][]{walker1981,berner1983,Berner1997,Sleep2001b,Abbot2012,Foley2016_review}. Plate tectonics plays a critical role in facilitating the carbonate-silicate cycle on Earth, and has thus often been posited to be a necessary component for sustaining a long-lived habitable climate on any rocky planet \citep[e.g.][]{Kasting2003}.  
	
However, whether plate tectonics is actually required for the carbonate-silicate cycle to operate and stabilize a planet's climate is unclear \citep[e.g.][]{Lenardic2016}. Stagnant lid planets still experience volcanism, which can release a significant amount of CO$_2$ to the atmosphere, as estimated in previous studies for Mars \citep[e.g.][]{Pollack1987,ONeill2007_mars,Grott2011}, a hypothetical stagnant lid Earth \citep{Tosi2017}, and super-Earths \citep{Noack2017,Dorn2018}. Moreover, burial of carbonated crust under lava flows could lead to metamorphic decarbonation of this crust, providing an additional CO$_2$ source to the atmosphere, or recycling of surface CO$_2$ back into the mantle \citep[e.g.][]{Pollack1987,Lindy2007}. \cite{Foley2018_stag} modeled the evolution of CO$_2$ outgassing rates for Earth-sized stagnant lid planets, considering both volcanic and metamorphic outgassing, and found that habitable climates could last for 2-5 Gyrs, depending on the amount of radiogenic elements the planet acquires during its formation, and the amount of CO$_2$ in the mantle and surface reservoirs. Specifically there is an optimal range of CO$_2$ budgets that allow for long-lived habitable climates. Too low a CO$_2$ budget and even with significant volcanism, outgassing rates are too low to maintain a warm climate. In this case a snowball state would prevail, possibly with limit cycles where long-lived snowball states are punctuated by short-lived hothouse climates \citep{Kadoya2014,Menou2015,Haqq2016,Batalha2016}. Too high a CO$_2$ budget and outgassing rates overwhelm weathering, causing an uninhabitably hot climate to form. The higher the abundance of heat producing elements the planet acquires, the longer volcanism lasts, and, when CO$_2$ budgets are within the optimum range for habitability, the longer habitable climates last as well. \cite{Valencia2018} obtained similar results considering planets that are tidally heated, in which case volcanism and outgassing could last as long as the planet's orbit leads to significant tides.     

While an important step in assessing the habitability of stagnant lid planets, \cite{Foley2018_stag} only tracks CO$_2$ outgassing rates, and assumes that outgassing rates must be at least 10-100 \% of Earth's present day degassing rate in order to prevent a cold, snowball climate, and must be below the global weathering supply limit in order to prevent an extreme hothouse. Actual climate evolution is not tracked, and doing so might change the estimated range of CO$_2$ budgets or radiogenic heating budgets that lead to long-lived habitability. Another issue not addressed in \cite{Foley2018_stag} is whether a stagnant lid planet could ever recover from a snowball climate, should one develop due to some perturbation in outgassing rate or weathering rate. The thermal evolution models used in this study, and in \cite{Foley2018_stag}, track the evolution of global average outgassing rates over billion year timescales, but can not capture short timescale fluctuations in outgassing (on the order of 10s of millions of years) that could result from the time dependent nature of mantle convection. As such, perturbations that could trigger a snowball climate, even when long term average outgassing rates should be high enough to keep the climate warm, are possible. In this study a model for weathering and climate evolution is included with the thermal evolution and CO$_2$ outgassing model of \cite{Foley2018_stag}, to determine how climate actually evolves on stagnant lid planets, re-assess the conditions that lead to long-lived habitable climates, and assess whether snowball climates can be recovered from. 

The models presented here make some important assumptions. First, only planets with a similar size and composition to Earth are considered, as the key material parameters governing mantle convection are highly uncertain for large super-Earth planets, where pressures in the mantle are far higher than experiments can reach \citep[e.g.][]{Karato2010,Stamenkovic2011}, and for planets with significantly different compositions than Earth. I also assume all melt produced is buoyant and will rise to the surface, as the melt generation depth is shallower than typical estimates for the silicate melt density crossover, as shown in \cite{Foley2018_stag}. Stagnant lid super-Earth planets, larger than 3 or 4 Earth masses, may actually experience limited outgassing, due to silicate melts being generated at such high pressures that the melts are denser than the surrounding mantle \citep{Noack2017,Dorn2018}. Another important assumption is that a constant solar luminosity, equal to the solar constant for Earth, is assumed. This is primarily done for simplicity, as the focus of this study is on how the thermal evolution of stagnant lid planets' mantles controls their climate evolution and climate stability. 
Adding stellar evolution would include an additional independent variable in the model, overcomplicating the results and their interpretation. 

\section{Theory}
\label{sec:theory}

\subsection{Thermal evolution model}
\label{sec:thermal_evol}

The thermal evolution model follows after \cite{Foley2018_stag}, which gives a thorough explanation of the model. As a result, only a brief overview of the model equations will be provided here. Pure internal heating is assumed, as this greatly simplifies the modeling and provides the most conservative estimate for the lifetime of outgassing and habitable climates. The thermal evolution models also include heat loss due to melting; all melt generated is assumed to contribute to mantle cooling and the growth of the crust, while only a fraction of the melt is assumed to erupt at the surface. These assumptions result in the following equation for the evolution of mantle temperature \citep[e.g.][]{Stevenson1983,Hauck2002,Reese2007,Fraeman2010,Morschhauser2011,Driscoll2014,Foley2018_stag}: 
\begin{equation}
\eqlbl{thermal_evol}
V_{man} \rho c_p \frac{dT_p}{dt} = Q_{man} - A_{man} F_{man} - f_m \rho_m \left (c_p \Delta T_m + L_m \right) ,
\end{equation} 
where $V_{man}$ is the volume of convecting mantle beneath the stagnant lid, $\rho$ is the bulk density of the mantle, $c_p$ is the heat capacity, $T_p$ is the potential temperature of the upper mantle, $t$ is time, $Q_{man}$ is the radiogenic heating rate in the mantle, $A_{man}$ is the surface area of the top of the convecting mantle, $F_{man}$ is the heat flux from the convecting mantle into the base of the stagnant lid, $f_m$ is {the} volumetric melt production rate, $\rho_m$ is the density of mantle melt, $L_m$ is the latent heat of fusion of the mantle, and $\Delta T_m$ is the temperature difference between the melt erupted at the surface and the surface temperature. The volume of the convecting mantle is $V_{man} = (4/3) \pi ((R_p-\delta)^3 - R_c^3)$, where $R_p$ is the planetary radius, $R_c$ is the core radius, and $\delta$ is the thickness of the stagnant lid. The surface area of the convecting mantle is then $A_{man} = 4 \pi (R_p-\delta)^2$. As Earth-sized planets are considered in this study, $R_p$ is set equal to Earth's radius and $R_c$ equal to Earth's core radius.

The thickness of the stagnant lid, $\delta$, evolves with time as \citep[e.g.][]{Schubert1979,Spohn1991}, 
\begin{equation}
\eqlbl{delta1}
\rho c_p (T_p - T_l) \frac{d \delta}{dt} = -F_{man} - k \frac{\partial T}{\partial z} \Bigr |_{z=R_p-\delta},
\end{equation}
where $T_l$ is the temperature at the base of the lid, $k$ is the thermal conductivity, and $z$ is height above the planet's center. As volcanism is assumed to cause heat loss from the mantle directly to the surface, melt heat loss does not appear in \eqref{delta1}. The mantle heat flux, $F_{man}$, and lid base temperature, $T_l$, are calculated from stagnant lid convection scaling laws as \citep{Reese1998,Reese1999,Solomatov2000b,korenaga2009}: 
\begin{equation}
\eqlbl{heat_flux}
F_{man} = \frac{c_1 k(T_p - T_{bot})}{d} \theta^{-4/3} Ra_i^{1/3}
\end{equation}
and 
\begin{equation}
\eqlbl{T_l}
T_l = T_p - \frac{a_{rh} R T_p^2}{E_v} .
\end{equation}
In \eqref{heat_flux} \& \eqref{T_l}, $c_1$ and $a_{rh}$ are constants, with assumed values $c_1 = 0.5$ and $a_{rh} = 2.5$. $T_{bot}$ is the temperature at the bottom of the oceans, which are assumed here to cover most of the planet's surface and thus set the top boundary condition for the convecting mantle (see \S \ref{sec:carbon_cycle}), $d=R_p-R_c$ is the whole mantle thickness, $\theta$ is the Frank-Kamenetskii parameter, defined as $\theta = E_v (T_p - T_{bot})/(RT_p^2)$, where $E_v$ is the activation energy for mantle viscosity and $R$ is the gas constant, and $Ra_i$ is the internal Rayleigh number. The internal Rayleigh number is defined as $Ra_i = \rho g \alpha (T_p - T_{bot}) d^3/(\kappa \mu_i)$, where $g$ is gravitational acceleration, $\alpha$ is thermal expansivity, $\kappa$ is thermal diffusivity, and $\mu_i$ is the interior mantle viscosity, defined at temperature $T_p$. Mantle viscosity follows the temperature-dependent viscosity law $\mu_i = \mu_n \exp{(E_v/(RT_p))}$, where $\mu_n$ is a constant; $E_v = 300$ kJ mol$^{-1}$ is the baseline activation energy used in this study \citep{karato1993}, and test cases using $E_v = 200$ kJ mol$^{-1}$ and $E_v = 400$ kJ mol$^{-1}$ are shown in \S \ref{sec:sensitivity}. {A baseline reference viscosity of $\mu_r = 2 \times 10^{20}$ Pa s is assumed.} The reference viscosity is defined as $\mu_r = \mu_n \exp{(E_v/(RT_r))}$, where $T_r = 1623$ K is Earth's present day mantle temperature; {this then defines the constant $\mu_n$ as $\mu_n = \mu_r  \exp{(-E_v/(RT_r))}$. With $E_v = 300$ kJ mol$^{-1}$ $\mu_n = 4 \times 10^{10}$ Pa s} (see Tables \ref{tab_param} \& \ref{tab_var} for a complete list of parameters and variables in the model).   

\begin{table}
\small
\begin{threeparttable}
\caption{Model Parameters and Assumed Values}
\label{tab_param}
\begin{tabular}{c c c c}
\hline
Parameter & Meaning & Assumed baseline value & Equation \\
\hline
$\rho$ & Mantle density & 4000 kg m$^{-3}$ & \eqref{thermal_evol} \\
$c_p$ & Heat capacity & 1250 J kg$^{-1}$ K$^{-1}$ & \eqref{thermal_evol} \\
$\rho_m$ & Melt density & 2800 kg m$^{-3}$ & \eqref{thermal_evol} \\
$L_m$ & Latent heat & 600 $\times 10^3$ J kg$^{-1}$ & \eqref{thermal_evol} \\
$R_p$ & Planet radius & 6378.1 km & below \eqref{thermal_evol} \\
$R_c$ & Core radius & 3488.1 km & below \eqref{thermal_evol} \\
$k$ & Thermal conductivity & 5 W m$^{-1}$ K$^{-1}$ & \eqref{delta1} \\
$c_1$ & Scaling law constant for $\delta$ & 0.5 & \eqref{heat_flux} \\
$d$ & Whole mantle thickness & 2890 km & \eqref{heat_flux} \\
$T_s$ & Surface temperature & 273 K & below \eqref{heat_flux} \\
$a_{rh}$ & Constant for rheological temperature scale & 2.5 & \eqref{T_l} \\
$E_v$ & Activation energy for viscosity & 300 kJ mol$^{-1}$ & \eqref{T_l} \\
$R$ & Universal gas constant & 8.314 J K$^{-1}$ mol$^{-1}$ & \eqref{T_l} \\
$g$ & Gravity & 9.8 m s$^{-2}$ & below \eqref{T_l} \\
$\kappa$ & Thermal diffusivity & $10^{-6}$ m$^2$ s$^{-1}$ & below \eqref{T_l} \\
$\alpha$ & Thermal expansivity & $3 \times 10^{-5}$ K$^{-1}$ & below \eqref{T_l} \\
$\mu_n$ & Viscosity pre-exponential factor & $4 \times 10^{10}$ Pa s & below \eqref{T_l} \\
$\mu_r$ & Reference viscosity & $2 \times 10^{20}$ Pa s & below \eqref{T_l} \\
$\gamma_{mantle}$ & Mantle adiabatic temperature gradient & $10^{-8}$ K Pa$^{-1}$ & \eqref{p_melt} \\
$\rho_l$ & Lithosphere density & 3300 kg m$^{-3}$ & \eqref{pf} \\
$(d \phi /d P)_S$ & Pressure derivative of melt fraction & $1.5 \times 10^{-10}$ Pa$^{-1}$ & \eqref{phi} \\
$\gamma$ & Effective adiabatic temperature gradient of melt & $6.766 \times 10^{-4}$ K km$^{-1}$ & below \eqref{fm} \\
$c_2$ & Scaling law constant for $v$ & 0.05 & \eqref{vel1} \\
$D$ & Distribution coefficient & 0.002 & \eqref{Q_crust} \\
$\tau_{rad}$ & Radioactive decay constant & 2.94 Gyrs & \eqref{Q_crust} \\
$A$ & Decarbonation temperature constant & $3.125 \times 10^{-3}$ K m$^{-1}$ & \eqref{T_dcarb} \\
$B$ & Decarbonation temperature constant & $835.5$ K & \eqref{T_dcarb} \\
{$D_{CO_2}$} & {Distribution coefficient for CO$_2$} & {$10^{-4}$} & {\eqref{Fd} }\\
$E_{fsw}$ & Seafloor weathering activation energy & $75000$ J mol$^{-1}$ & \eqref{Fweather} \\
$\beta$ & Sensitivity of seafloor weathering to eruption rate & $1/2$ & \eqref{Fweather} \\
$F^*_{fsw}$ & Modern day Earth seafloor weathering rate & $0.5$ Tmol yr$^{-1}$ & \eqref{Fweather} \\
$T^*_{bot}$ & Modern day Earth ocean bottom temperature & $274$ K & \eqref{Fweather} \\
{$\eta^*$} & Modern day Earth eruption rate & $23 \times 10^9$ km$^3$ yr$^{-1}$ & \eqref{Fweather} \\
$\epsilon$ & Eruption efficiency & $0.1$ & \eqref{Fweather} \\
$\chi$ & Weathering demand of basalt & 5.8 mol kg$^{-1}$ & \eqref{fws} \\
$T^*_s$ & Present day Earth surface temperature & 285 K & \eqref{surf_temp} \\
$p\mathrm{CO_2}^*$ & Partial pressure of atmospheric CO$_2$ & 30 Pa & \eqref{surf_temp} \\
$\bar{m}_{CO_2}$ & Molar mass of CO$_2$ & 0.044 kg mol$^{-1}$ & below \eqref{surf_temp} \\
$A_{surf}$ & Surface area of the planet & $5.11 \times 10^{14}$ m$^2$ & below \eqref{surf_temp} \\
$h_{\phi}$ & Atmosphere-ocean CO$_2$ partitioning parameter & $10^{20}$ mol & \eqref{phi_aoc} \\
$T_{i}$ & Initial mantle potential temperature & 2000 K & \S \ref{sec:init} \\
\end{tabular} 
\end{threeparttable}    
\end{table}


The temperature gradient at the base of the stagnant lid, as needed in \eqref{delta1}, is calculated assuming one-dimensional heat conduction in steady-state, with constant radiogenic heating rates in the crust and mantle. The crustal radiogenic heating rate is defined as $x_c = Q_{crust}/V_{crust}$, where $x_c$ is the crustal heat production rate per unit volume, $Q_{crust}$ is the total radiogenic heating rate in the crust, and $V_{crust}$ is the volume of the crust. The mantle radiogenic heating rate $x_m = Q_{man}/(V_{man}+V_{lid})$, where $x_m$ is the mantle heat production rate per unit volume, $V_{lid} = (4/3)\pi((R_p-\delta_c)^3-(R_p-\delta)^3)$ is the volume of the sub-crustal stagnant lid, and $\delta_c$ is the thickness of the crust. The temperature profile, and hence lid base temperature gradient, can be solved for analytically in this case, giving 
\begin{equation}
\eqlbl{flux_lid}
{ -k \frac{\partial T}{\partial z} \Bigr |_{z=R_p-\delta} = \frac{k(T_l-T_c)}{\delta-\delta_c} - \frac{x_m(\delta-\delta_c)}{2} }
\end{equation}      
when $\delta > \delta_c$, and 
\begin{equation}
\eqlbl{flux_crust}
-k \frac{\partial T}{\partial z} \Bigr |_{z=R_p-\delta} = -\frac{x_c \delta_c}{2} + \frac{k(T_l-T_{bot})}{\delta_c} 
\end{equation} 
when $\delta = \delta_c$. The temperature at the base of the crust, $T_c$, is given by 
\begin{equation}
\eqlbl{T_c}
{T_c = \frac{T_{bot}(\delta-\delta_c)+T_l\delta_c}{\delta} + \frac{x_c \delta_c^2(\delta-\delta_c) + x_m \delta_c(\delta-\delta_c)^2}{2 k \delta}.}
\end{equation} 
The thermal conductivity, $k$, is assumed constant; the difference between crustal and mantle thermal conductivity is ignored for simplicity. As explained in \cite{Foley2018_stag}, \eqref{flux_crust} is used when $T_c$ is within one degree Kelvin of $T_l$, which is equivalent to switching when the difference between $\delta_c$ and $\delta$ is negligibly small. The crustal temperature profile is also calculated assuming that advection is negligible. \cite{Foley2018_stag} showed that advection is small compared to diffusion, though not necessarily negligible. As a result of ignoring advection, the lid thickness may be slightly underestimated in our models, and hence conductive cooling overestimated. Advection may also suppress metamorphic decarbonation of the crust, so a case where decarbonation is assumed to not occur is presented in \S \ref{sec:sensitivity}.     

\subsection{Melting and crustal evolution}
\label{sec:melting}

The melt production rate, $f_m$, is calculated as in \cite{Foley2018_stag}, which is based on \cite{Fraeman2010}. The pressure where melting begins, $P_i$, is 
\begin{equation}
\eqlbl{p_melt}
P_i = \frac{T_p - 1423}{120 \times 10^{-9} - \gamma_{mantle}} , 
\end{equation}
where $\gamma_{mantle}$ is the average adiabatic temperature gradient in the upper mantle ($\approx 10^{-8}$ K Pa$^{-1}$). Assuming melting stops at the base of the stagnant lid, the pressure where melting stops is
\begin{equation}
\eqlbl{pf}
P_f = \rho_l g \delta ,
\end{equation} 
where $\rho_l$ is the average density of the crust and lithosphere (assumed to be $\rho_l = 3300$ kg m$^{-3}$). The melt fraction, $\phi$, is 
\begin{equation}
\eqlbl{phi}
\phi = \frac{P_i - P_f}{2} \left(\frac{d\phi}{dP} \right)_S,
\end{equation}
where $(d\phi/dP)_S \approx 1.5 \times 10^{-10}$ Pa$^{-1}$. 

The full melt production rate is then given by assuming a cylindrical region of upwelling mantle enters the melting zone beneath the stagnant lid \citep[as in, e.g.][]{Reese1998}. A derivation is given in \cite{Foley2018_stag}, which results in 
\begin{equation}
\eqlbl{fm}
f_m = 17.8 \pi R_p v (d_m - \delta) \phi,
\end{equation}
where $v$ is the convective velocity, $d_m$ is the depth where melting begins ($d_m = P_i/(\rho_lg$)), and assuming that $d \approx 0.45 R_p$ as on Earth. The exact form of the melt production rate was found to not significantly impact the results in \cite{Foley2018_stag}. The temperature difference between melt erupted at the surface and the surface temperature is $\Delta T_m = T_p - P_i \gamma - T_s$ where $T_s$ is the (atmospheric) surface temperature of the planet and $\gamma$ was found to be $\gamma = 6.766 \times 10^{-4}$ K km$^{-1}$ based on values in \cite{Driscoll2014}. The velocity, $v$, is calculated from scaling laws for internally heated stagnant lid convection \citep{Reese1998,Reese1999,Solomatov2000b,korenaga2009} as
\begin{equation}
\eqlbl{vel1}
v = c_2 \frac{\kappa}{d} \left(\frac{Ra_i}{\theta} \right)^{2/3} ,
\end{equation}
where $c_2$ is a constant.  

All melt generated is assumed to contribute to crust growth, and any crust buried to depths below $\delta$ is assumed to founder into the mantle, such that the lithospheric thickness always follows \eqref{delta1}. With the above assumptions, the evolution of crustal volume, $V_{crust}$, is
\begin{equation}
\eqlbl{crust2}
\frac{d V_{crust}}{dt} = f_m - \left(f_m-4\pi(R_p-\delta)^2\textrm{min}\left(0,\frac{d\delta}{dt}\right) \right)(\tanh{((\delta_c-\delta)20)}+1).
\end{equation} 
The hyperbolic tangent function allows the crustal loss rate to go to zero when $\delta_c < \delta$, and equal the crustal growth rate plus the lid thinning rate when $\delta_c = \delta$. The term $4\pi(R_p-\delta)^2\textrm{min}(0,d\delta/dt)$ gives the rate at which the volume of stagnant lid is lost when $\delta$ is shrinking, and is 0 when $\delta$ is growing. The crustal thickness, $\delta_c$, is calculated as
\begin{equation}
\eqlbl{delta_c}
\delta_c = R_p -\left( R_p^3 - \frac{3 V_{crust}}{4 \pi} \right)^{1/3},
\end{equation}
assuming that crustal thickness is constant spatially across the planet. 

Heat producing elements are incompatible and partition preferentially into the melt. Accumulated fractional melting is assumed in this study \citep[e.g.][]{Fraeman2010,Morschhauser2011}, resulting in the following equations for the evolution of the crustal heat production rate, $Q_{crust}$, 
\begin{equation}
\begin{split}
\eqlbl{Q_crust}
\frac{d Q_{crust}}{dt} & = \frac{x_m f_m}{\phi} [1 - (1-\phi)^{1/D} ] - \\ & x_c \left(f_m-4\pi(R_p-\delta)^2\textrm{min}\left(0,\frac{d\delta}{dt}\right) \right)(\tanh{((\delta_c-\delta)20)}+1) - \frac{Q_{crust}}{\tau_{rad}}
\end{split}
\end{equation} 
and mantle heat production rate, $Q_{man}$,
\begin{equation}
\begin{split}
\eqlbl{Q_man}
\frac{d Q_{man}}{dt} & = x_c \left(f_m-4\pi(R_p-\delta)^2\textrm{min}\left(0,\frac{d\delta}{dt}\right) \right)(\tanh{((\delta_c-\delta)20)}+1) - \\ & \frac{x_m f_m}{\phi} [1 - (1-\phi)^{1/D} ] - \frac{Q_{man}}{\tau_{rad}}.
\end{split}
\end{equation} 
The distribution coefficient, $D$, is assumed to have a value $D=0.002$ \citep{Hart1993,Hauri1994}. A constant decay constant for both the crust and mantle, $\tau_{rad} \approx 2.94$ Gyrs, is used based on an average of the four major radiogenic isotopes powering Earth's mantle, $^{238}$U, $^{235}$U, $^{232}$Th, and $^{40}$K \citep{Driscoll2014}.

\begin{table}
\small
\begin{threeparttable}
\caption{Model Variables}
\label{tab_var}
\begin{tabular}{c c c}
\hline
Variable & Meaning {(units)} & Equation \\
\hline
$T_p$ & Mantle potential temperature {(K)} & \eqref{thermal_evol} \\
$t$ & Time {(s)} &\eqref{thermal_evol} \\
$V_{man}$ & Volume of convecting mantle {(m$^3$)} & \eqref{thermal_evol} \\
$A_{man}$ & Surface are of top of convecting mantle {(m$^2$)} & \eqref{thermal_evol} \\
$Q_{man}$ & Radiogenic heat production in the mantle {(W)} & \eqref{thermal_evol} and \eqref{Q_man} \\
$F_{man}$ & Mantle convective heat flux {(W m$^{-2}$)} & \eqref{thermal_evol} \\
$f_m$ & Volumetric melt production rate {(m$^3$ s$^{-1}$)} & \eqref{thermal_evol} \\
$\Delta T_m$ & Temperature difference between erupted melt and $T_{s}$ {(K)} &  \eqref{thermal_evol} and below \eqref{fm} \\
$\delta$ & Thickness of stagnant lid {(m)} & \eqref{delta1} and \eqref{pf} \\
$z$ & Height above planet's center {(km)} & \eqref{delta1} \\
$T_l$ & Temperature at base of stagnant lid {(K)} & \eqref{delta1} and \eqref{T_l} \\
$T_{bot}$ & Temperature at the bottom of the ocean (K) & below \eqref{heat_flux} and \eqref{bot_temp} \\
$\theta$ & Frank-Kamenetskii parameter {(unit-less)} & \eqref{heat_flux} \\
$Ra_i$ & Internal Rayleigh number {(unit-less)}  & \eqref{heat_flux} \\
$\mu_i$ & Interior mantle viscosity {(Pa s)}  & below \eqref{T_l} \\
$x_c$ & Concentration of heat producing elements in the crust {(W m$^{-3}$)} &  below \eqref{T_l} and \eqref{flux_crust} \\
$Q_{crust}$ & Radiogenic heat production in the crust {(W)} & below \eqref{T_l} and \eqref{Q_crust} \\
$V_{crust}$ & Volume of crust {(m$^3$)} & below \eqref{T_l} and \eqref{crust2} \\
$x_m$ & Concentration of heat producing elements in the crust {(W m$^{-3}$)} & below \eqref{T_l} and \eqref{flux_lid} \\
$V_{lid}$ & Volume of subcrustal stagnant lid {(m$^3$)} & below \eqref{T_l} \\
$\delta_c$ & Thickness of crust {(m)} & \eqref{T_c} and \eqref{delta_c} \\
$T_c$ & Temperature at the base of the crust {(K)} & \eqref{T_c} \\
$P_i$ & Pressure where melting begins {(Pa)} & \eqref{p_melt} \\
$P_f$ & Pressure where melting ceases {(Pa)} & \eqref{pf} \\
$\phi$ & Melt fraction {(unit-less)} & \eqref{phi} \\
$d_m$ & Depth where melting begins {(m)} & \eqref{fm} \\
$v$ & Convective velocity {(m s$^{-1}$)} & \eqref{vel1} \\
$T_{decarb}$ & Temperature of metamorphic decarbonation {(K)} & \eqref{T_dcarb} \\
$\delta_{carb}$ & Decarbonation depth {(m)} & \eqref{d_carb} \\
$R_{man}$ & Mantle carbon reservoir {(mol)} & \eqref{Fd} and \eqref{rman} \\
$F_d$ & Mantle degassing flux {(mol s$^{-1}$)} & \eqref{Fd} \\
$R_{crust}$ & Crustal carbon reservoir {(mol)} & \eqref{f_meta} and \eqref{rcrust} \\
$F_{meta}$ & Metamorphic degassing flux {(mol s$^{-1}$)} & \eqref{f_meta} \\
$V_{carb}$ & Volume of carbonated crust {(m$^{3}$)} & \eqref{f_meta} \\
$F_{weather}$ & Seafloor weathering flux {(mol s$^{-1}$)} & \eqref{Fweather} \\
$F_{sl}$ & Supply limited weathering flux {(mol s$^{-1}$)} & \eqref{fws} \\
$T_{s}$ & Surface temperature (K) & \eqref{surf_temp} \\
$p\mathrm{CO_2}$ & Partial pressure of atmospheric CO$_2$ (Pa) & \eqref{surf_temp} \\
$\phi_{aoc}$ & Fraction of atmosphere-ocean CO$_2$ in atmosphere (unit-less) & \eqref{phi_aoc} \\
$R_{aoc}$ & Atmosphere-ocean carbon reservoir (mol) & \eqref{raoc} \\
$C_{tot}$ & Total carbon budget {(mol)} & \S \ref{sec:init} \\
$Q_{0}$ & Initial radiogenic heat production {(W)} & \S \ref{sec:init} \\
\end{tabular} 
\end{threeparttable}    
\end{table}

\subsection{Carbon cycle and climate} 
\label{sec:carbon_cycle}

Cycling of CO$_2$ is tracked in a similar fashion to \cite{Foley2018_stag}, though with some important differences: in this study CO$_2$ in the atmosphere and oceans is explicitly tracked, surface temperature is calculated from a simple climate model, and weathering rates are calculated to link outgassing, atmospheric CO$_2$, and climate. Carbon is assumed to outgas to the atmosphere-ocean system via both mantle volcanism and metamorphic decarbonation of buried crust. Once in the atmosphere-ocean system weathering removes CO$_2$ to the crust, where subsequent volcanism buries the crust until it either experiences metamorphic decarbonation, or the carbonated crust founders back into the mantle. The atmosphere-ocean system is assumed to always be in equilibrium, and CO$_2$ is partitioned between the atmosphere and ocean accordingly (see \eqref{phi_aoc}).  

The temperature-pressure conditions where carbonate-bearing metabasalt experiences decarbonation are determined from the P-T phase diagram shown in \cite{Foley2018_stag} \citep[see also][]{Staudigel1989,Kerrick2001b}.  With a simple linear fit to the 0.25 wt. \% CO$_2$ contour, the temperature at which decarbonation occurs is
\begin{equation}
\eqlbl{T_dcarb}
T_{decarb} = A (R_p-z) + B
\end{equation} 
where $A=3.125 \times 10^{-3}$ K m$^{-1}$, $B=835.5$ K, $T_{decarb}$ is temperature in Kelvins, and pressure is converted into depth by assuming a lithospheric density of $\rho_l = 3300$ kg m$^{-3}$. The depth where decarbonation occurs, $\delta_{carb}$, is then given by \citep[see][for details]{Foley2018_stag} 
\begin{equation}
\begin{split}
\eqlbl{d_carb}
\delta_{carb} = & \frac{\delta_c}{2} + \frac{k(T_c-T_{bot})}{\delta_c x_c} - \frac{Ak}{x_c} - \\ & \frac{k}{x_c} \sqrt{\left(\frac{x_c \delta_c}{2k} + \frac{T_c-T_{bot}}{\delta_c} - A \right)^2 + \frac{2 x_c}{k}(T_{bot} - B)}.
\end{split}
\end{equation} 


There are three carbon fluxes in the model: the mantle degassing flux due to volcanism, $F_d$, the metamorphic degassing flux due to decarbonation of the crust, $F_{meta}$, and the weathering flux, $F_{weather}$. The mantle degassing flux is given by the product of the concentration of CO$_2$ in mantle melt and the mantle melt production rate,
\begin{equation}
\eqlbl{Fd}
F_d = \frac{f_m R_{man} [1 - (1-\phi)^{1/D_{CO_2}} ]}{\phi (V_{man}+V_{lid})}, 
\end{equation}
where $R_{man}$ is the mantle CO$_2$ reservoir, in mols of CO$_2$. The formulation for $F_d$ takes into account the fact that CO$_2$ is an incompatible element, like heat producing elements, with a distribution coefficient of $D_{CO_2} = 10^{-4}$ \cite[e.g.][]{Hauri2006}. The metamorphic degassing flux is given by the product of the concentration of CO$_2$ in the crust and the volumetric melt production rate, as  
\begin{equation}
\eqlbl{f_meta}
F_{meta} = \frac{R_{crust}f_m}{2 V_{carb}}(\tanh{((\delta_c-\delta_{carb})20)}+1) 
\end{equation}
where $R_{crust}$ is the crustal CO$_2$ reservoir in mols, $V_{carb}$ is the volume of carbonated crust, $V_{carb} = (4/3) \pi (R_p^3 - (R_p - \delta_{carb})^3)$, and, as in \eqref{crust2}, a hyperbolic tangent function is used as a mathematically convenient way to parameterize a step-function like change in crustal CO$_2$ around the decarbonation depth. Here a factor of 1/2 is included, so that the metamorphic degassing flux is $R_{crust} f_m/(V_{carb})$ when $\delta_c > \delta_{carb}$ and zero when $\delta_c < \delta_{carb}$. {Note that in this formulation metamorphic outgassing does not begin until the crust first grows deep enough to reach the decarbonation depth.} Metamorphic CO$_2$ outgassing is assumed to be complete and to occur uniformly across the planet's surface, as crustal production is assumed to be spatially uniform. Incomplete metamorphic outgassing was found to not significantly impact the results in \cite{Foley2018_stag}. {The above formulation for the metamorphic outgassing flux also assumes that the crust acts as a single reservoir with instantaneous mixing, when in reality mixing between different crustal layers will not occur. However, assuming a mixed crustal reservoir does not significantly impact the results, as shown in Appendix \ref{sec:appendix}, and greatly simplifies the model.}   

The rate of CO$_2$ drawdown from the atmosphere-ocean system is calculated by assuming that seafloor weathering is the dominant weathering process. Seafloor weathering is assumed because stagnant lid planets are likely to have crusts that are predominantly basaltic \citep{Breuer2007}, as there is no subduction to produce continental crust as on Earth. Furthermore, without continents, planets will not have well defined ocean basins, so oceans may be more wide spread across the planets' surface. Of course no actual constraints on topography and the size of oceans for real stagnant lid exoplanets are available, so the assumption here of seafloor weathering being the primary weathering process may not be correct. However, it is a simple first order assumption to make based on our knowledge of stagnant lid versus plate tectonic planets. The weathering flux includes a supply limit, as in \cite{Foley2015_cc} \& \cite{Foley2018_stag}, and is temperature-dependent as in \cite{KT2017}. The plate speed dependence of seafloor weathering from \cite{KT2017} is replaced with a dependence on the melt eruption flux, as it is ultimately the crustal production rate that matters for seafloor weathering. However the dependence of seafloor weathering on ocean pH is ignored, as it is significantly weaker than the temperature dependence. 

The CO$_2$ drawdown rate is thus 
\begin{equation}
\eqlbl{Fweather}
{
F_{weather} = F_{sl} \left \{ 1- \exp{\left[-\frac{F_{sfw}^*}{F_{sl}}\exp{\left(-\left(\frac{E_{sfw}}{R} \right) \left(\frac{1}{T_{bot}} - \frac{1}{T_{bot}^*} \right) \right)} \left(\frac{\epsilon f_m}{\eta^*} \right)^{\beta}\right ]}  \right \}
}
\end{equation} 
where $F_{sl}$ is the supply limit to seafloor weathering, $F_{sfw}^*$ is the modern day Earth seafloor weathering rate and is a constant, $E_{sfw}$ is the activation energy for seafloor weathering, $T_{bot}^*$ is the modern day Earth ocean bottom temperature, $\epsilon$ is the eruption efficiency, or the fraction of melt produced that erupts at the surface {(and thus $\epsilon f_m$ is the eruption rate)}, and {$\eta^*$} is the eruption rate at modern Earth mid-ocean ridges. To prevent numerical convergence problems { associated with very small atmosphere-ocean carbon reservoir sizes}, $F_{weather}$ is set to 0 when surface temperatures fall below 260 K. {In reality weathering could continue in a snowball state, in particular if exchange between atmosphere and sub-ice ocean occurs \citep[e.g.][]{Tajika2008}, so climate could cool even beyond the 260 K cutoff imposed here. However, surface temperatures typically only fall to 260 K as volcanism and outgassing are waning, at which point habitable climates are assumed to end anyway in this study (see \S \ref{sec:baseline_results}).} The baseline values of $F_{sfw}^*$ and $E_{sfw}$ are taken from \citep{KT2017}, the baseline value for $\epsilon$ is $\epsilon = 0.1$ \citep{Foley2018_stag}, and for {$\eta^*$} is {$\eta^* = 23 \times 10^9$} km$^3$ yr$^{-1}$ \citep{Crisp1984} (see Table \ref{tab_param}).

The supply limit to weathering is \citep{Foley2015_cc,Foley2018_stag} 
\begin{equation}
\eqlbl{fws}
F_{sl} = \epsilon f_m \chi \rho_l,
\end{equation}   
where $\chi$ is the weathering demand for complete carbonation of basalt, analogous to the formulation in \cite{Kump2018} for continental crust. The weathering demand is calculated from the mass fraction of reactable oxides in basalt: CaO, MgO, FeO, K$_2$O, and Na$_2$O. Mass fractions from \cite{Gale2013} are used, which gives 5.8 moles of oxide per kg of basalt. Hence $\chi = 5.8$ mol kg$^{-1}$ of net CO$_2$ drawdown by completely carbonating basalt. The temperature at the base of the ocean is related to the surface (atmospheric) temperature of the planet as \citep{KT2017} 
\begin{equation}
\eqlbl{bot_temp}
T_{bot} = \max{(270,1.02T_s - 16.7)};
\end{equation}
the temperature is not allowed to fall below 270 K at the bottom of the ocean because even during a cold climate the bottom of the ocean is not expected to freeze. $T_{bot}^*$ is $T_{bot}^* = 1.02T_s^* -16.7$, where $T_s^*=285$ K is Earth's present day surface temperature. The surface temperature is calculated from the simple parameterization given in \cite{KT2017}, 
\begin{equation}
\eqlbl{surf_temp}
T_s = T_s^* + 5.6\left( \frac{\ln{(p\mathrm{CO}_2/p\mathrm{CO}_2^*)}}{\ln{(2)}} \right) ,
\end{equation} 
where $p\mathrm{CO}_2$ is the partial pressure of atmospheric CO$_2$, and $p\mathrm{CO}_2^*$ is the pre-industrial $p\mathrm{CO}_2$ for the Earth ($p\mathrm{CO}_2^* = 30$ Pa). The partial pressure of atmospheric CO$_2$ is defined as $p\mathrm{CO}_2 = R_{atm}\bar{m}_{CO_2} g/A_{surf}$, where $R_{atm}$ is the size of the atmospheric CO$_2$ reservoir in mol, $\bar{m}_{CO_2}$ is the molar mass of CO$_2$, and $A_{surf}$ is the surface area of the planet. Finally, $R_{atm}$ is related to the ocean plus atmosphere CO$_2$ reservoir, $R_{aoc}$, as $R_{atm} = \phi_{aoc} R_{aoc}$, where 
\begin{equation}
\eqlbl{phi_aoc}
\phi_{aoc} = \frac{0.78 R_{aoc}}{R_{aoc} + h_{\phi}}
\end{equation}
as in \cite{Mills2011}; here $h_{\phi} = 10^{20}$ mol. Note that the climate parameterization used in this study can not capture a snowball, as this would significantly change the parameterization for surface temperature as a function of atmospheric CO$_2$. However, the ability of planets to recover from snowball states is assessed in this study based on outgassing rates (see \S \ref{sec:snowball}). 

When $\delta_{carb} < \delta$, crust decarbonates before being recycled into the mantle. The evolution of the combined atmosphere-ocean CO$_2$ reservoir is then given by  \citep[e.g.][]{TajikaMatsui1992,Franck1999,Sleep2001b,Driscoll2013,Foley2015_cc}
\begin{equation}
\eqlbl{raoc}
\frac{dR_{aoc}}{dt} = F_d + F_{meta} - F_{weather} ,
\end{equation}
the evolution of the mantle carbon reservoir, $R_{man}$, by
\begin{equation}
\eqlbl{rman}
\frac{dR_{man}}{dt} = -F_d ,
\end{equation} 
and the evolution of the crustal carbon reservoir, $R_{crust}$, by
\begin{equation}
\eqlbl{rcrust}
\frac{dR_{crust}}{dt} = F_{weather} - F_{meta} .
\end{equation} 
As CO$_2$ is not returned to the mantle when crustal rocks decarbonate before reaching the base of the stagnant lid, volcanic degassing of the mantle causes a net transfer of carbon from the mantle to the crust over the lifetime of a stagnant lid planet in this case.  

When $\delta_c < \delta_{carb}$, the following equations for $R_{aoc}$, $R_{man}$, and $R_{crust}$ are used: 
\begin{equation}
\eqlbl{raoc2}
\frac{dR_{aoc}}{dt} = F_d - F_{weather}
\end{equation}
\begin{equation}
\begin{split}
\eqlbl{rman2}
\frac{dR_{man}}{dt} & = \frac{R_{crust}}{V_{crust}}\left(f_m-4\pi(R_p-\delta)^2\textrm{min}\left(0,\frac{d\delta}{dt}\right) \right)(\tanh{((\delta_c-\delta)20)}+1) - F_d.
\end{split}
\end{equation}
\begin{equation}
\begin{split}
\eqlbl{rcrust2}
\frac{dR_{crust}}{dt} & = F_{weather} - \frac{R_{crust}}{V_{crust}}\left(f_m-4\pi(R_p-\delta)^2\textrm{min}\left(0,\frac{d\delta}{dt}\right) \right)(\tanh{((\delta_c-\delta)20)}+1)
\end{split}
\end{equation} 
In this case, crust buried beneath the stagnant lid base will founder into the mantle, returning surface carbon to the mantle. However, in practice $\delta_{carb}$ is usually found to be less than $\delta_c$ while volcanism is active, so models typically experience crustal decarbonation and do not recycle CO$_2$ from the crust back into the mantle. 

\section{Model setup}
\subsection{Initial conditions}
\label{sec:init}

The above equations for the evolution of mantle temperature, crust volume, CO$_2$ cycling, and surface temperature are solved simultaneously as a set of coupled ordinary differential equations. In order to solve the system initial conditions and some additional constraints must be specified. First, two key input quantities that are varied in the models are the initial radiogenic heating rate, $Q_0$, and the total CO$_2$ budget of the surface and mantle reservoirs, $C_{tot}$; as CO$_2$ among these reservoirs is conserved, $C_{tot} = R_{crust} + R_{man} + R_{aoc}$. All models start with a chosen initial heat production rate, $Q_0$, that is entirely within the mantle. $Q_0$ is varied over a range of 5-250 TW, encompassing typical estimates for the Earth's initial mantle radiogenic heating rate, which span $\approx 70-100$ TW when present day heating rates in the mantle and continental crust are combined and extrapolated back in time \citep[e.g.][]{korenaga2006}. Models also start with an initial crust volume of zero, that is no crust is present; as volcanism causes the crust to grow, heat producing elements are then partitioned into the crust. Models are also run for a wide range of $C_{tot} = 3 \times 10^{18} - 3 \times 10^{24}$ mol, bracketing the Earth's estimated value of $\sim 10^{22}$ mol \citep{Sleep2001b}.   

In the case of CO$_2$, initial conditions are important as the distribution of CO$_2$ between the different reservoirs determines the planet's climate. Two end member limits are considered here: an initially hot climate (or ``hot start" scenario) and an initially cool climate (or ``cold start" scenario). In the initially hot climate, 99.99 \% of the CO$_2$ in the planet's surface and mantle reservoirs is placed in the atmosphere-ocean system, with the rest residing in the mantle. This initial condition is a possible result of magma ocean solidification, which is predicted to result in significant outgassing from the mantle to the surface, and hence an initially CO$_2$ rich atmosphere \citep{Zahnle2007,Lindy2008}. In the cold start scenario, the initial surface temperature is held fixed at 273 K, and CO$_2$ partitioned between reservoirs accordingly. That is, the amount of CO$_2$ required to produce $T_s = 273$ K is placed in the atmosphere-ocean system, and the remainder is placed in the mantle. Colder initial surface temperatures are not considered, since the climate model in this study can not explicitly capture snowball climates. The initial temperature of the mantle, $T_i$, is typically set at $T_i = 2000$ K, a reasonable value for a recently solidified silicate mantle \citep[e.g.][]{Abe1997,Solomatov2000,Foley2014_initiation}. However, a range of initial mantle temperatures from $T_i = 1500-2200$ K is tested (see \S \ref{sec:sensitivity}). Finally, the initial stagnant lid thickness is calculated from \eqref{delta1} in steady-state at the specified initial mantle temperature, surface temperature, and heat production rate. 

\subsection{Calculating recovery from snowball states}
\label{sec:snowball}

A major goal of this paper is determining whether stagnant lid planets could recover from snowball episodes. To determine this, two snowball scenarios are considered: a ``hard" snowball, where there is no exchange between atmosphere and ocean and no weathering during the snowball episode, and a ``soft" snowball where atmosphere and ocean are assumed to maintain equilibrium. In the soft snowball, the partitioning fraction of CO$_2$ between atmosphere and ocean, $\phi_{aoc}$, is modified by setting $h_{\phi} = 5.3 \times 10^{20}$ mol, as in \cite{Mills2011}, which results in more CO$_2$ being partitioned into the ocean, and less in the atmosphere, during the snowball phase than during a non-snowball climate. I assume that in order to recover from the snowball phase, the partial pressure of atmospheric CO$_2$ must reach $2.5 \times 10^4$ Pa of CO$_2$, as in \cite{Mills2011} and consistent with estimates from pervious studies \citep[e.g.][]{Caldeira1992b,Hoffman2002}. For a soft snowball, this constraint on $p\mathrm{CO}_2$ means $\approx 1.6 \times 10^{20}$ mol of CO$_2$ are needed in the atmosphere-ocean reservoir to melt the ice layer, and for a hard snowball $\approx 3 \times 10^{19}$ mol of CO$_2$ are needed in the atmosphere. 

For a hard snowball, only volcanism releasing CO$_2$ above the ice layer can contribute to warming the climate. Metamorphic outgassing is likely confined to low lying regions that are beneath the ice layer, so I assume that all CO$_2$ released by crustal decarbonation is released to the oceans, and therefore does not contribute to warming the climate during a hard snowball state. Mantle volcanism, however, is assumed to occur at volcanoes that reach above the ice layer, and thus release of mantle CO$_2$ is assumed to contribute to warming the climate. In reality some of the CO$_2$ released by mantle volcanism may also be confined to the sub-ice oceans, so the assumptions here present an optimistic scenario for recovering from a hard snowball state. In the soft snowball, both metamorphic and volcanic outgassing contribute to warming the climate, as equilibrium between atmosphere and ocean is assumed. 

The thermal evolution models produce a time history of outgassing and volcanism; this history is then used to determine if a hypothetical snowball state could be recovered from. For a hard snowball, the outgassing rate due to mantle volcanism, $F_d$, is integrated over time as: 
\begin{equation}
\int_{t'}^{t_{end}} F_d(t) dt
\end{equation}
 where $t'$ is the time when a hypothetical hard snowball state is assumed to occur, and $t_{end}$ is 10 Ga, the time when the thermal evolution models are stopped. Once the integrated value of CO$_2$ released reaches the threshold needed to melt the ice layer, $p\mathrm{CO}_2 = 2.5 \times 10^4$ Pa, then the snowball is considered to be recovered from. The integration is performed at each timestep of the model output, {and the latest time during model evolution at which a hard snowball state can be recovered from is recorded. That is, for any time after this there is not enough CO$_2$ that can be outgassed from the mantle, before volcanism dies out, to melt the ice layer. Thus,} the age at which the planet can no longer recover from a hard snowball due to insufficient degassing, is determined.   
 
As with the hard snowball case, in the soft snowball case I determine the { planet age after which} a soft snowball state can {not} be recovered from. For each timestep in the model output, corresponding to a time $t'$, I calculate whether metamorphic and volcanic outgassing can bring the system out of a hypothetical soft snowball state, by increasing the atmosphere-ocean CO$_2$ reservoir size to at least $1.6 \times 10^{20}$ mol. In the case of a soft snowball, carbon cycling between the crust, mantle, and atmosphere needs to be tracked as weathering will still be active, drawing CO$_2$ out of the combined ocean-atmosphere system, and the crustal CO$_2$ reservoir will be changing over time. Thus at each timestep, $t'$, equations \eqref{raoc} \& \eqref{rcrust} are solved forward in time to see if enough CO$_2$ can accumulate in the atmosphere to melt the ice. The degassing flux, $F_d(t)$, and the volcanism rate, $f_m(t)$, are taken from the thermal evolution model outputs. The initial conditions for the atmosphere-ocean and crustal CO$_2$ reservoirs for the hypothetical snowball state, here labeled $R_{aoc}^*(t')$ and $R_{crust}^*(t')$, are determined as follows. I assume that in order to trigger the snowball state, atmospheric CO$_2$ levels must have fallen to 15 Pa \citep[e.g.][]{Donna2004,Mills2011}. The amount of CO$_2$ in the ocean is then determined by \eqref{phi_aoc} with $h_{\phi} = 5.3 \times 10^{20}$, hence also setting $R_{aoc}^*(t')$. The crustal CO$_2$ abundance at the start of the snowball state, $R_{crust}^*(t')$, is given by $R_{crust}^*(t') = R_{crust}(t') + R_{aoc}(t') - R_{aoc}^*(t')$, where $R_{crust}(t')$ and $R_{aoc}(t')$ are the amounts of CO$_2$ in the crust and atmosphere-ocean, respectively, at time $t'$ of the thermal evolution model output (i.e. before the hypothetical soft snowball climate develops). In other words, I assume that enough CO$_2$ has been removed from the atmosphere-ocean system in order to trigger a snowball state, and that this CO$_2$ has been weathered out into the crust. 
 
Some models are run where metamorphic decarbonation of the crust is assumed to not occur, and all crustal carbon recycles into the mantle. Calculating snowball recovery in this case is slightly different than in the cases where decarbonation is allowed to occur. For a soft snowball state, { the same procedure as described in the preceding paragraph is followed, except equations \eqref{raoc2}-\eqref{rcrust2} are integrated, with $f_m(t)$ as output from the thermal evolution model results used in the integration. Initial conditions for the crust, mantle, and atmosphere-ocean carbon reservoirs are the same as described above.} For a hard snowball, {\eqref{rman2} \& \eqref{rcrust2} are integrated forward in time, with the seafloor weathering flux set to zero, so that depletion of the crustal and mantle carbon reservoirs are taken into account.} That is, as there is no exchange of CO$_2$ from the atmosphere to the ocean and hence back into the crust {during a hard snowball}, the {crustal CO$_2$ reservoir will be depleted by foundering into the mantle, and the} mantle CO$_2$ reservoir will be depleted by volcanic degassing.   

\section{Results}
\subsection{Baseline results and influence of initial conditions}
\label{sec:baseline_results}

\begin{figure}
\includegraphics[width=0.5\textwidth]{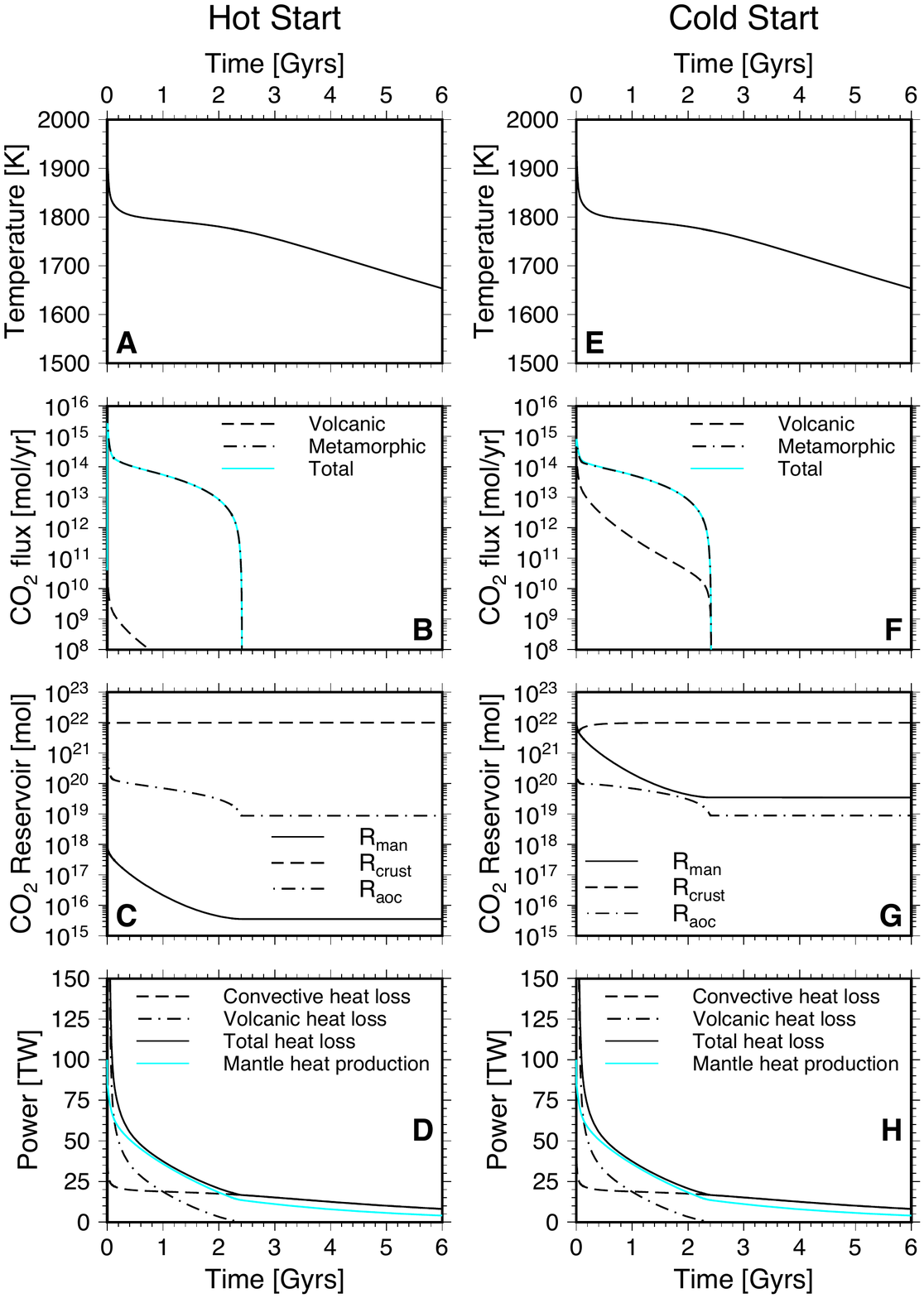}
\caption{\label {fig:model_evolution}  Time evolution for models with $Q_0=100$ TW and $C_{tot}=10^{22}$ mol, and the rest of the parameters as given in Table \ref{tab_param}. Both a model with an initially hot climate and one with an initially cold climate are shown. Time evolution of mantle temperature (A \& E), mantle, metamorphic, and total CO$_2$ outgassing rates (B \& F), mantle, crustal, and combined atmosphere-ocean reservoir sizes (C \& G), and mantle heat loss and heat production (D \& H) are shown.}
\end{figure}

\begin{figure}
\includegraphics[width=0.5\textwidth]{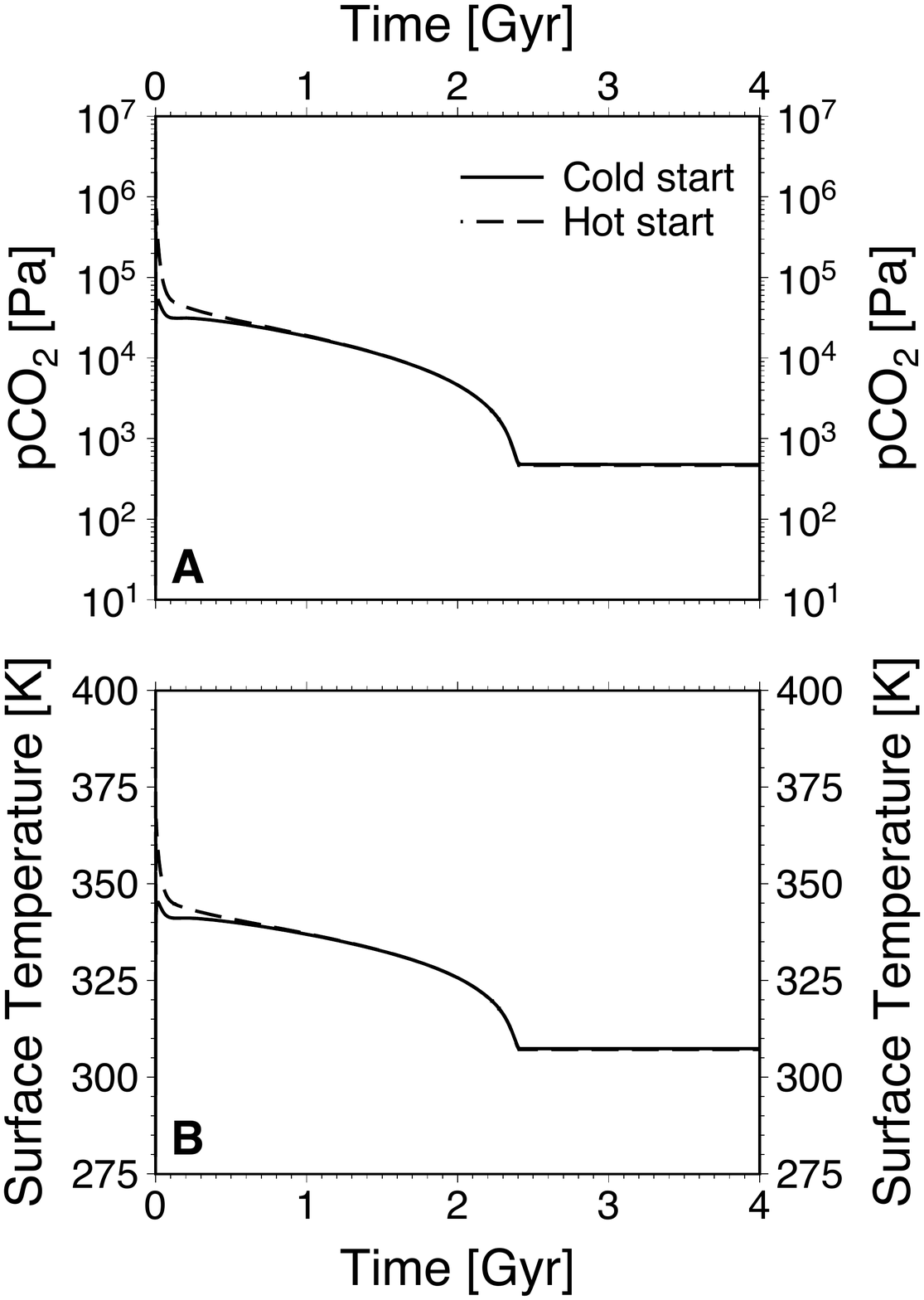}
\caption{\label {fig:climate_evolution}  Time evolution of atmospheric $p\mathrm{CO}_2$ (A) and surface temperature (B), for the hot and cold start models presented in Figure \ref{fig:model_evolution}.}
\end{figure}

Example thermal evolution model results are given in Figures \ref{fig:model_evolution} \& \ref{fig:climate_evolution}. Specifically two models with identical model parameters, but different initial distributions of CO$_2$ between mantle and atmosphere-ocean, are shown to compare how these different initial conditions influence the overall model evolution. The parameter values are as given in Table \ref{tab_param}, which are considered the baseline model parameters for this study. The different initial climate states have no influence on the evolution of mantle temperature or mantle heat loss (Figure \ref{fig:model_evolution}A \& E, D \& H), because the difference in surface temperature between the two cases does not significantly influence stagnant lid convection, and climate states for the two cases converge early in planetary evolution (Figure \ref{fig:climate_evolution}). However, the volcanic CO$_2$ outgassing rate and evolution of the mantle CO$_2$ reservoir are strongly affected by the initial condition (Figure \ref{fig:model_evolution}B \& F, C \& G). With a hot start, the mantle is depleted in CO$_2$, and CO$_2$ is never recycled back into the mantle over time due to metamorphic decarbonation of the crust. As a result, $R_{man}$ and $F_d$ are very low throughout model evolution. With an initially cold climate, mantle outgassing is initially significant, but the mantle is quickly depleted of CO$_2$ and metamorphic outgassing becomes the primary source of atmospheric CO$_2$. The total outgassing rate and amount of CO$_2$ in the atmosphere-ocean reservoirs follow nearly identical evolutionary paths for either initial condition. As a result, the climate evolution is not strongly affected by the initial climate state (Figure \ref{fig:climate_evolution}). For both a hot and cold start, atmospheric CO$_2$ and surface temperature evolutionary paths are nearly identical, after a short initial adjustment phase. During this adjustment phase outgassing leads to a rapid warming for the initially cold climate case, and weathering leads to rapid cooling of the initially hot climate. The initial climate condition has no influence on when mantle volcanism ends, at which point outgassing ceases and a cold climate is expected to develop, as explained below. The insensitivity of climate evolution to the two end member initial conditions used in this study is also confirmed by the more extensive set of results presented below.  

The thermal evolution models are used to calculate how long temperate climates, with surface temperatures between 273 K and 400 K, last, for a range of CO$_2$ budgets ($C_{tot}$) and initial radiogenic heating rates ($Q_0$). The longevity of habitable climates is only calculated for the time period when volcanism is active, as climate evolution without volcanism is uncertain. In the model it is assumed that weathering is entirely inactive when volcanism is inactive, as without volcanism there is no supply of fresh rock to the surface. However, even with no volcanism weathering can still take place, as long as there are still fresh, unaltered minerals in the weathering zone that groundwater can reach. Thus climate would likely cool over time without volcanism, until either the climate has cooled to a permanent snowball state or weathering ceases due to complete carbonation of the crust. Note that in Figure \ref{fig:climate_evolution} climate is still temperate when volcanism ends; however in reality climate would likely cool further due to continued weathering and the absence of outgassing. In this study I assume that planets will always cool to a snowball state when volcanism ends, which is a conservative assumption; habitable climates may persist after volcanism has ended (or exist before it begins, in cases where the mantle is initially cool), so the estimates presented here give the minimum lifetimes for habitable climates. Habitable lifetime is thus calculated as the minimum of the time when volcanism ends and the time when temperatures fall outside of the 273-400 K range, minus the maximum of the time when volcanism begins and surface temperatures first enter the 273-400 K habitable range.  

\begin{figure}
\includegraphics[width=1\textwidth]{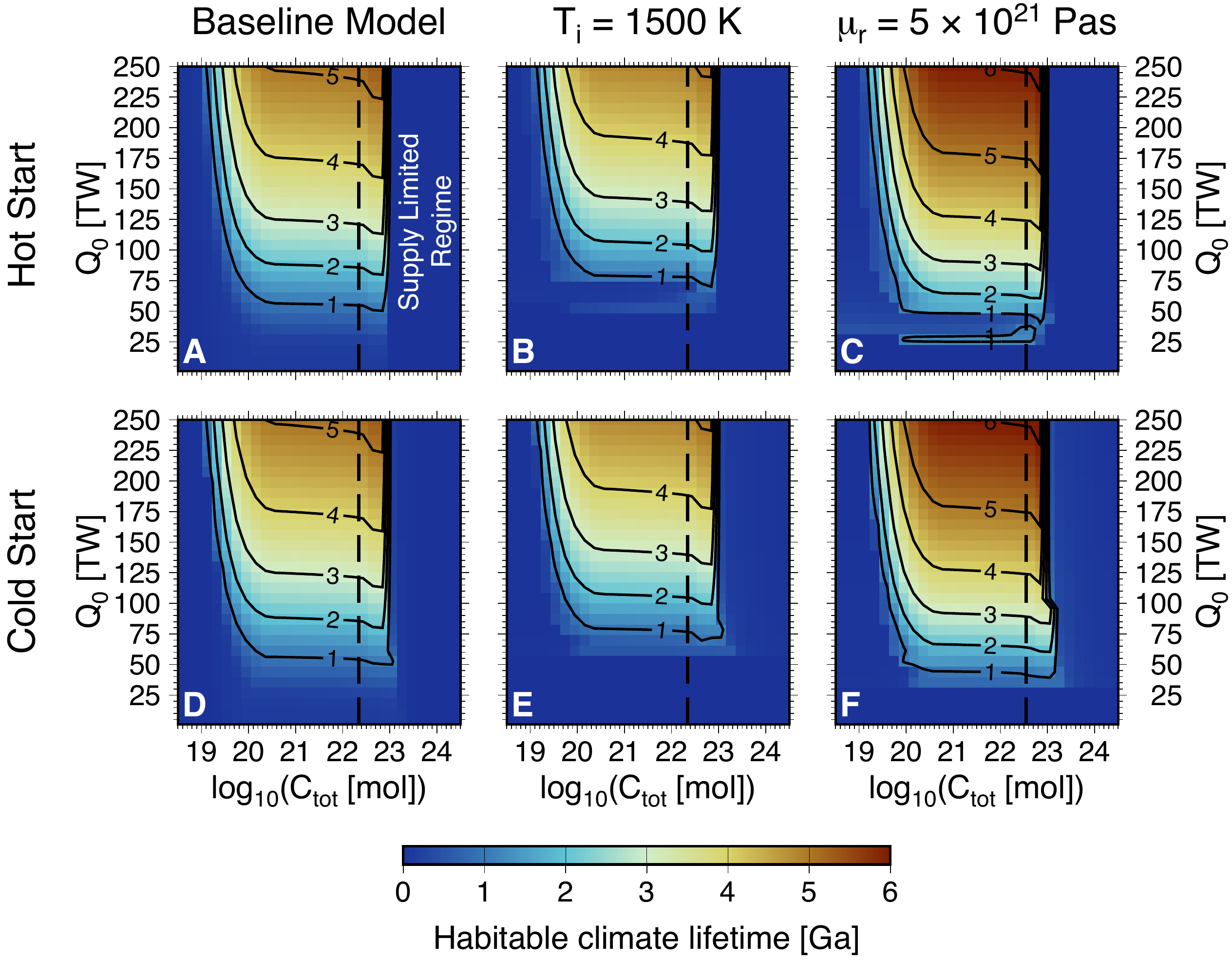}
\caption{\label {fig:habtime_compare}  Lifetime of habitable climates for stagnant lid planets as a function of initial mantle heat production rate, $Q_0$, and total CO$_2$ budget of the mantle and surface reservoirs, $C_{tot}$. Models using baseline parameters (see Table \ref{tab_param}) for both a hot and cold start scenario are shown in Figures \ref{fig:habtime_compare}A \& D, respectively; with $T_i =1500$ K in Figures \ref{fig:habtime_compare}B \& E; and with $\mu_r = 5 \times 10^{21}$ Pa s in Figures \ref{fig:habtime_compare}C \& F. The dashed line marks the onset of supply limited weathering; weathering is supply limited to the right of the dashed line. }
\end{figure} 

The lifetimes of habitable climates as a function of $Q_0$ and $C_{tot}$ are shown for both hot and cold start scenarios (Figure \ref{fig:habtime_compare}). At very low CO$_2$ budgets of $\sim 10^{19}$ mol or less, habitable climates are not possible, as outgassing rates are not sufficient to prevent surface temperatures from immediately dropping below the freezing point of water. That is, even when most of the planet's CO$_2$ has been outgassed into the atmosphere-ocean reservoir, there is insufficient warming to keep the planet above freezing. With increasing $C_{tot}$, habitable climate lifetime increases, until CO$_2$ budgets are so high that planets would enter the supply limited weathering regime. Here habitable climates are still possible for a narrow range of $C_{tot}$, where even with supply limited weathering the resulting surface temperature is still less than $400$ K. However, as $C_{tot}$ continues to increase, planets reach a point where climates are always too hot to support life as we know it. Thus there is an optimum range of $C_{tot}$ for habitability, from $~10^{19}-10^{23}$ mol. 

Initial mantle heat production rate also plays a critical role in habitable climate lifetime, by dictating how long volcanism and outgassing can last. For the baseline model (Figures \ref{fig:habtime_compare}A \& D), $Q_0 \gtrapprox 50$ TW is needed for habitable climates to last at least 1 Ga. With increasing $Q_0$, the habitable lifetime increases in turn, as long as $C_{tot}$ is within the range where habitable climates are possible; this is because higher rates of radiogenic heat production prolong volcanism and outgassing. With a lower initial mantle temperature of $T_i = 1500$ K (\ref{fig:habtime_compare}B \& E), the mantle is initially too cold for significant volcanism; thus radiogenic heating must warm the mantle up before volcanism and outgassing can begin. As a result, a higher $Q_0$ is needed before habitable climates lasting 1 Ga or more can be maintained, and habitable lifetimes are generally lower than in the baseline model, where the initial mantle temperature is larger. When the mantle reference viscosity is larger than the baseline value (\ref{fig:habtime_compare}C \& F), habitable lifetimes are longer because the mantle cools more slowly, and hence outgassing lasts longer.     

A key feature of Figure \ref{fig:habtime_compare} is that the habitable lifetimes for hot and cold start scenarios closely match. For the baseline model, there is almost no difference between models with either an initially hot or initially cold climate. With $T_i=1500$ K or $\mu_r = 5 \times 10^{21}$ Pa s there are some noticeable differences between the two initial conditions {at low values of $Q_0$}, but overall habitable lifetimes are still remarkably consistent. {The differences are caused by the short lived nature of volcanism at low $Q_0$. For the hot climate initial condition, volcanism does not last long enough for metamorphic decarbonation to begin or for weathering and outgassing to come into balance; the band of $\sim$ 1 Ga habitable lifetimes at $Q_0 \approx 25$ TW for $\mu_r = 5 \times 10^{21}$ Pa s and at $Q_0 \approx 50$ TW for $T_i = 1500$ K are caused by volcanism beginning, but not lasting long enough to draw down the initially CO$_2$ rich atmosphere. A small window of temperate climates thus results. However in the cold start cases, higher rates of volcanism are needed to keep the climate warm, in light of weathering and the much lower initial atmospheric CO$_2$ content. Thus higher heat production rates are required before even short lived habitable climates can develop with a cold start initial condition. Outside of these limited regions of parameter space, however,} the initial condition does not have a significant influence on long term evolution of climate, outgassing, and mantle thermal evolution, as long as the initial climate state is not a snowball (see \S \ref{sec:recovery} \& \ref{sec:hysteresis}), or so hot that liquid water cannot exist and weathering cannot operate. A climate that starts off initially hot cools rapidly due to weathering, while one that starts cold warms due to outgassing, until both models forget their initial conditions and follow the same evolution. 

\subsection{Recovery from snowball states}
\label{sec:recovery}

\begin{figure}
\includegraphics[width=0.7\textwidth]{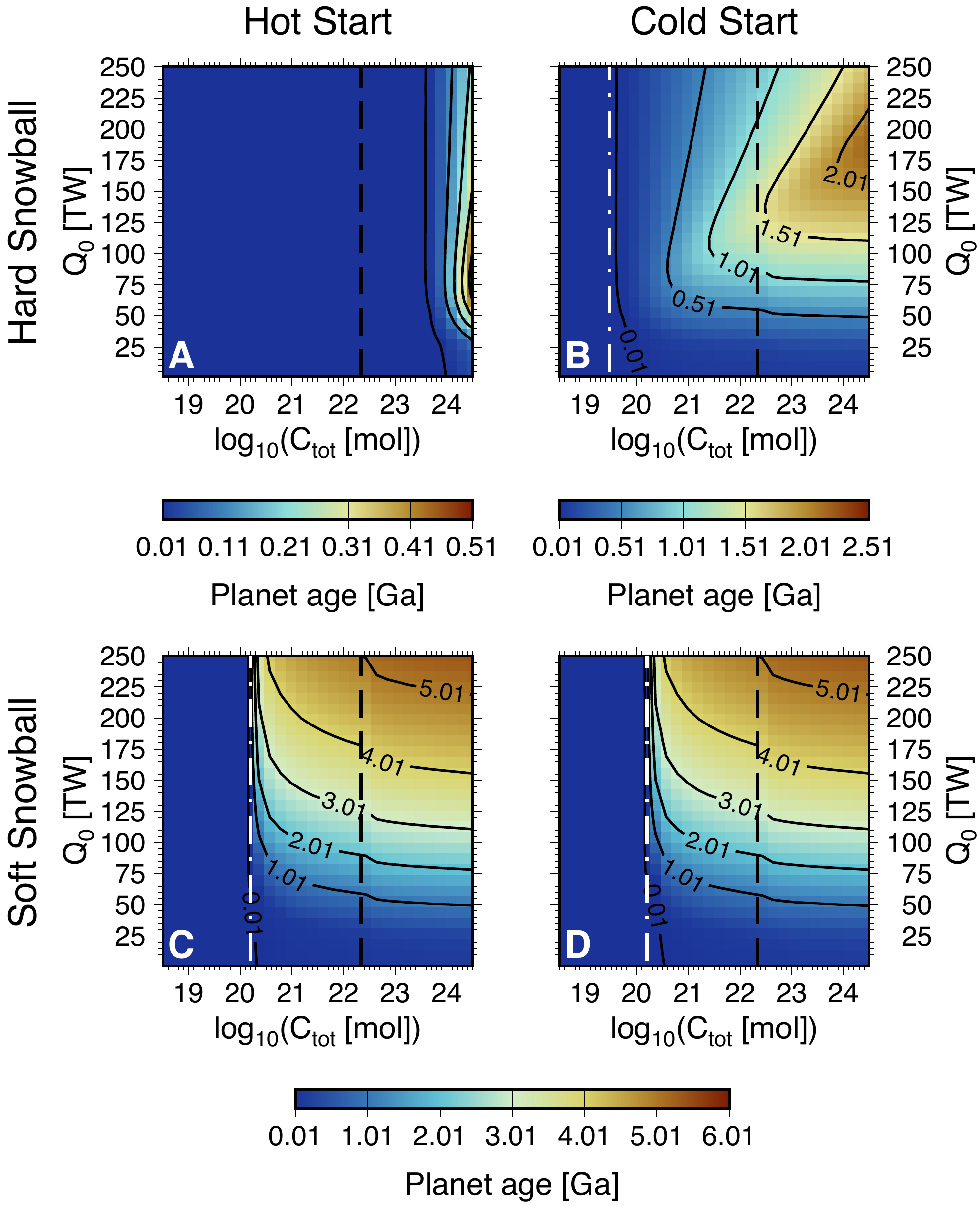}
\caption{\label {fig:recovertime_compare}  Planet age after which recovery from a snowball state is impossible for a hard snowball with a hot start (A), hard snowball with a cold start (B), soft snowball with a hot start (C), and soft snowball with a cold start (D). Weathering is supply limited to the right of the black dashed line, as in Figure \ref{fig:habtime_compare}. The dot-dashed white line in Figure \ref{fig:recovertime_compare}B at $C_{tot} = 3 \times 10^{19}$ mol gives the minimum value of $C_{tot}$ {needed for} an initial hard snowball climate {to} be recovered from, and in Figures \ref{fig:recovertime_compare}C \& D at $C_{tot} = 1.6 \times 10^{20}$ mol the minimum value of $C_{tot}$ {needed for} an initial soft snowball {to} be recovered from. }
\end{figure} 

However, different initial conditions do influence whether snowball states could be recovered from. As explained above in \S \ref{sec:snowball}, for each model, I calculate the {planet age after which} both a hard snowball and soft snowball state can {no longer} be recovered from. As outgassing is required to recover from a snowball state, these {ages} are related to the lifetime of volcanism; without CO$_2$ outgassing any snowball state would be permanent. The {planet ages after which recovery is impossible} for both hard and soft snowball states, with both hot and cold start initial conditions, are shown in Figure \ref{fig:recovertime_compare} for models with baseline parameter values. To recover from a hard snowball state, mantle volcanism must release at least $3 \times 10^{19}$ mol of CO$_2$ to the atmosphere. Thus, the mantle CO$_2$ content is critical for whether a hard snowball state can be recovered from. With an initially hot climate, most of the planet's CO$_2$ resides in the atmosphere, before being weathered out into the crust, and then cycling through the crust and returning to the atmosphere via metamorphic decarbonation. The mantle starts with only a small fraction of the planet's CO$_2$, and does not gain CO$_2$ from the surface reservoirs as decarbonation reactions do not allow for recycling of CO$_2$ back into the mantle. As a result, planets with an initially hot climate can not recover from a hard snowball state at any point during their evolution, unless the CO$_2$ budget is very large ($\gtrapprox 3 \times 10^{23}$ mol). Such large values of $C_{tot}$ are expected to result in supply limited weathering and uninhabitably hot climates, so planets within the range of $C_{tot}$ that allow for habitable climates would not be able to recover from hard snowball states, if the climate is initially hot.   

However, with an initially cold climate, more CO$_2$ initially resides in the planet's mantle, and recovery from a hard snowball state is possible. Even here, snowball recovery is only possible early in the planet's evolution, as the mantle quickly becomes depleted in CO$_2$ via outgassing. For example, with $C_{tot} = 10^{21}$ mol, recovery from a hard snowball state is only possible for the first 500 million years of a planet's history. Increasing $C_{tot}$ generally prolongs the time period over which recovery from a hard snowball is possible (Figure \ref{fig:recovertime_compare}B). Within the range of $C_{tot}$ where habitable climates are possible, planets can recover from hard snowball states when they are younger than up to 1-1.5 Gyrs. The initial radiogenic heating rate is also important and has two competing effects; increasing $Q_0$ allows volcanism to last longer, and mantle volcanism is essential for recovery from a hard snowball state. However, increasing $Q_0$ also increases the rate of volcanism, and hence depletion of mantle CO$_2$. Thus at low values of $Q_0$, increasing $Q_0$ allows snowball recovery to occur at later planet ages, because the end of volcanism is the limiting factor for whether a snowball state can be recovered from in this case. However, at even larger values of $Q_0$, increasing $Q_0$ lowers the age when snowball recovery becomes impossible, as the high rates of internal heating deplete the mantle of CO$_2$ more rapidly.   

An important end-member case is when a hard snowball state can never be recovered from, even if it occurs as a planet's initial surface condition. This case is important because it marks the point where a planet's initial climate state would dictate its later climate evolution; that is a planet that starts in a snowball would be stuck in this snowball for it's entire history, while a planet that starts with a hot or temperate climate would be able to avoid a snowball and potentially develop a habitable climate for at least some portion of its evolution. As $2.5 \times 10^4$ Pa of CO$_2$ in the atmosphere, corresponding to $3 \times 10^{19}$ mol, are required to recover from a snowball, a simple limit to when a hard snowball can never be recovered from is when $C_{tot} < 3 \times 10^{19}$ mol. This limit is plotted on Figure \ref{fig:recovertime_compare}B as a dot-dashed white line, and corresponds well to the model results where initial snowball states can not be recovered from. Thus, planets with $C_{tot} < 3 \times 10^{19}$ mol would be permanently stuck in a snowball state if they are initially frozen over, unless the luminosity of the star they orbit increases to such an extent that it melts the ice layer. 

Recovery from a soft snowball state is generally much easier than recovery from a hard snowball (Figures \ref{fig:recovertime_compare}C \& D). Soft snowball states can be recovered from for nearly the entire time period when volcanism is active. Thus, planets that manage to fall into a soft snowball state, due to some perturbation to their climate and carbon cycle, should be able to recover from this state throughout the time period when the models predict that habitable climates are possible. There is also no discernible difference between an initially hot climate state and initially cold climate state, in terms of their ability to recover from a soft snowball. As outgassing that leads to snowball recovery is dominated by metamorphic outgassing, the initial mantle CO$_2$ content is no longer important. However, even in the soft snowball case there is still a limit where an initially frozen climate would remain frozen throughout the planet's history, due to an inability to recover from the snowball climate. For a soft snowball, $1.6 \times 10^{20}$ mol of CO$_2$ must reside in the atmosphere-ocean system in order to melt the ice. Thus, planets with $C_{tot} < 1.6 \times 10^{20}$ mol will be unable to escape an initial snowball climate, and such a state would be permanent. The carbon budget below which snowball states are always permanent is actually higher for a soft snowball than hard snowball; this is because a significant portion of the outgassed CO$_2$ will be sequestered in the ocean during a soft snowball, where it does not warm the climate, and because seafloor weathering will remain active, removing CO$_2$ from the combined ocean-atmosphere system. 

\subsection{Hysteresis in stagnant lid planet habitability}
\label{sec:hysteresis}

\begin{figure}
\includegraphics[width=0.7\textwidth]{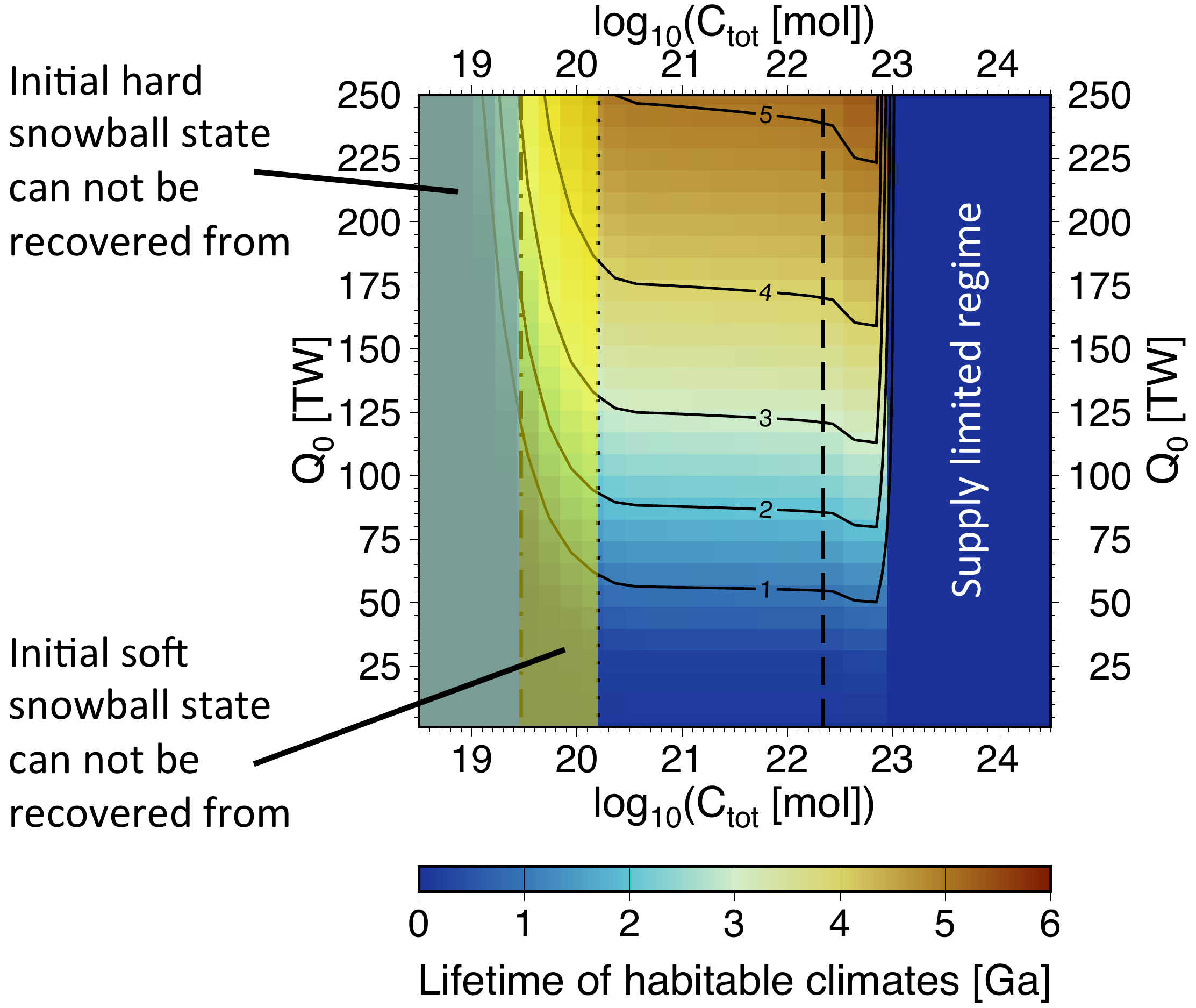}
\caption{\label {fig:summary}  Lifetime of habitable climates for the baseline model with an initially warm climate. {The dot-dashed line shows the minimum CO$_2$ budget, $C_{tot} = 3 \times 10^{19}$ mol, needed to recover from a hard snowball state, and the dotted line shows the minimum CO$_2$ budget, $C_{tot} = 1.6 \times 10^{20}$ mol, needed to recover from a soft snowball state. Planets with $C_{tot}$ lower than these limits could experience hysteresis, where with an initially frozen climate the planet would remain in this state and not be habitable, while with an initially warm climate habitable conditions could last as long as the calculated habitable lifetimes indicate. Weathering is supply limited when $C_{tot}$ is greater than the limit shown by the dashed line.}}
\end{figure} 

The inability of planets with low CO$_2$ budgets to recover from an initial snowball state means hysteresis is possible. Planets with an initially frozen climate may be stuck in this state permanently, and hence uninhabitable for surface life, while planets that start with an initially temperate or hot climate will be able to sustain habitable conditions, for at least a limited amount of time. Figure \ref{fig:summary} shows the baseline model predictions for the lifetime of habitable climates for planets with an initially hot climate, and denotes the regions of parameter space where initial snowball states can not be recovered from. Specifically, with $C_{tot} < 3 \times 10^{19}$ mol an initial hard snowball state could not be recovered from. However, at this $C_{tot}$ limit, most planets will not be capable of sustaining habitable climates regardless of the initial condition; when $C_{tot} < 10^{19}$ mol habitable climates can not be maintained for 1 Gyrs, even with $Q_0 = 250$ TW. Thus hysteresis based on initial condition is not a significant effect for a hard initial snowball. However, for a soft snowball hysteresis is more important, as a higher CO$_2$ budget of $1.6 \times 10^{20}$ mol is required to recover from this state. The $C_{tot} < 1.6 \times 10^{20}$ mol limit encompasses a significant portion of the region of $Q_0-C_{tot}$ space where habitable climates can be maintained. Another important aspect of hysteresis in stagnant lid planet climate evolution is that planets that fall into a snowball state at some later point during their evolution could become stuck in these states, as shown in Figure \ref{fig:recovertime_compare}. In particular planets that enter a hard snowball state are unlikely to recover, unless significant quantities of CO$_2$ still reside in the mantle, meaning planets one might otherwise predict to be habitable could end up stuck in an unrecoverable snowball, should some climate or carbon cycle perturbation kick them into this state during their evolution. Soft snowball states can generally be recovered from while volcanism is active, as long as $C_{tot} > 1.6 \times 10^{20}$ mol, so hysteresis during later planetary evolution is less likely in this case.   

\subsection{Sensitivity of model results}
\label{sec:sensitivity}

\begin{figure}
\includegraphics[width=1\textwidth]{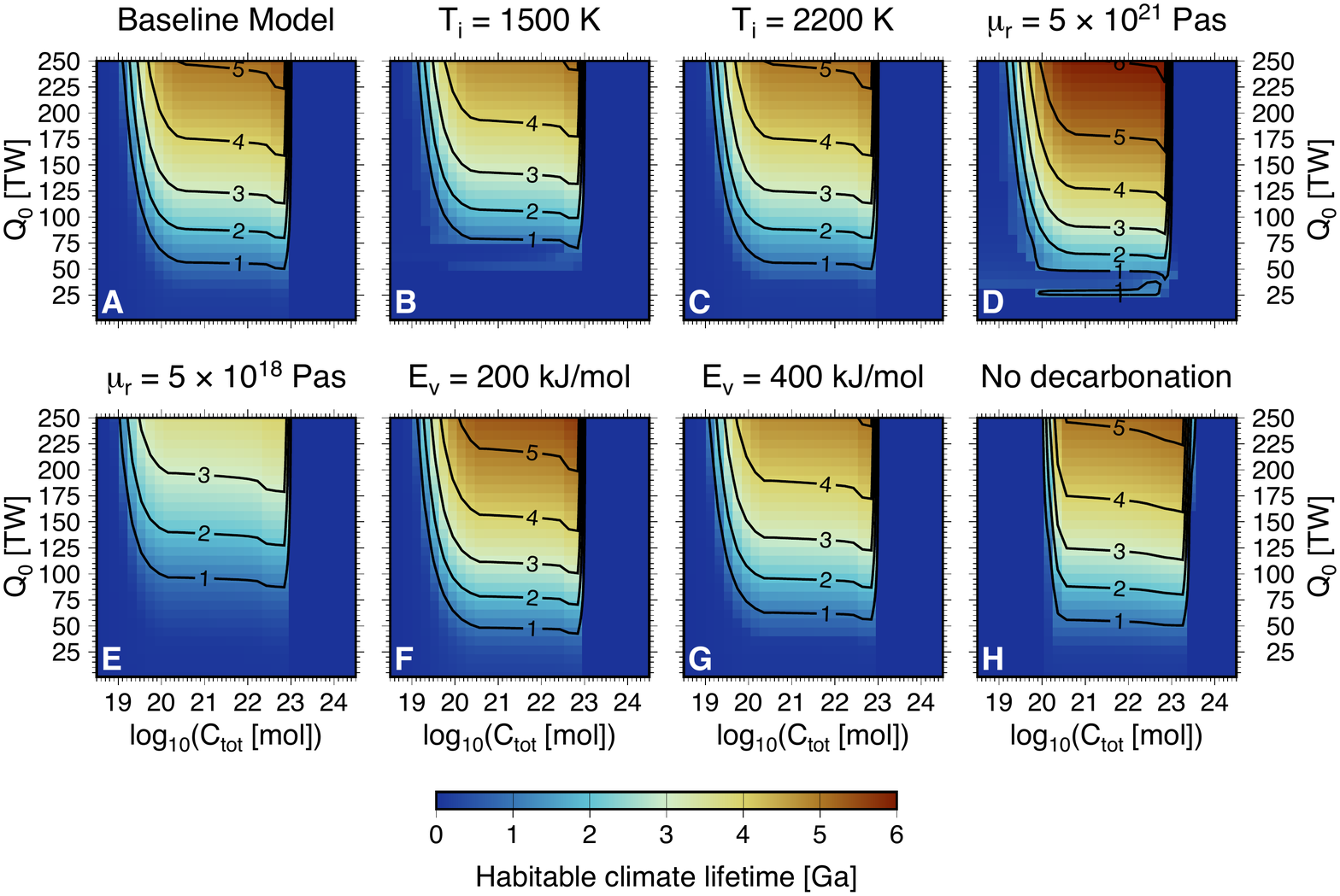}
\caption{\label {fig:habtime_sens}  Lifetime of habitable climates for planets with an initially hot climate and (A) baseline model parameters, (B) $T_i = 1500$ K, (C) $T_i = 2200$ K, (D) $\mu_r = 5 \times 10^{21}$ Pa s, (E) $\mu_r = 5 \times 10^{18}$ Pa s, (F) $E_v = 200$ kJ/mol, (G) $E_v = 400$ kJ/mol, (H) no crustal decarbonation. }
\end{figure} 

The overall model results are consistent when varying the initial mantle temperature, reference viscosity, viscosity activation energy, and when crustal decarbonation is ignored. However, there are some key differences worth discussing. Habitable lifetimes for a hot start initial condition are shown in Figure \ref{fig:habtime_sens}; only models with an initially hot climate are shown, as those with an initially cold climate show very similar habitable lifetimes. Decreasing the initial mantle temperature lowers the habitable climate lifetimes for all models, because the lower initial mantle temperature decreases the longevity of volcanism. Furthermore, higher values of $Q_0$ are required for habitable climates to last at least 1 Gyr, because the initial mantle temperature is not high enough for volcanism, and thus the mantle must heat before volcanism can commence (Figure \ref{fig:habtime_sens}B). With a higher initial mantle temperature of $T_i = 2200$ K, there is no discernible difference compared to the baseline model, as volcanism can begin immediately for all models with either $T_i = 2000$ K, as in the baseline models, or with $T_i = 2200$ K (Figure \ref{fig:habtime_sens}C). Changing the reference viscosity has an important influence on habitable climate lifetime; with a larger reference viscosity, habitable climates last longer due to a slower cooling rate of the mantle, while with a lower reference viscosity the opposite is true, and habitable climate lifetimes are everywhere shorter (Figure \ref{fig:habtime_sens}D \& E). The activation energy for viscosity influences habitable lifetimes in a similar manner. Lower values of $E_v$ prolong temperate climates by causing the mantle to cool more slowly, while a higher $E_v$ causes the mantle to cool more rapidly, and hence shortens the lifetime of temperate climates (Figure \ref{fig:habtime_sens}F \& G). Finally, when decarbonation of the crust is assumed to not occur, such that all crustal CO$_2$ recycles back into the mantle, higher total CO$_2$ budgets are needed for long-lived temperate climates (Figure \ref{fig:habtime_sens}H). Specifically, $C_{tot} > 10^{20}$ mol is needed for a habitable climate to last at least 1 Gyr, compared to $C_{tot} > 10^{19}$ mol when decarbonation of the crust is considered. However, temperate climates extend to larger $C_{tot}$ in the no decarbonation case, before outgassing overwhelms weathering and leads to uninhabitability hot climates. The reason for this behavior is the much larger volume of the mantle compared to the crust. The same total CO$_2$ budget will lead to overall lower outgassing rates when outgassing is primarily due to mantle volcanism rather than crustal decarbonation, as the large volume of the mantle dilutes the CO$_2$ concentration of rocks experiencing melting and outgassing. 

\begin{figure}
\includegraphics[width=1\textwidth]{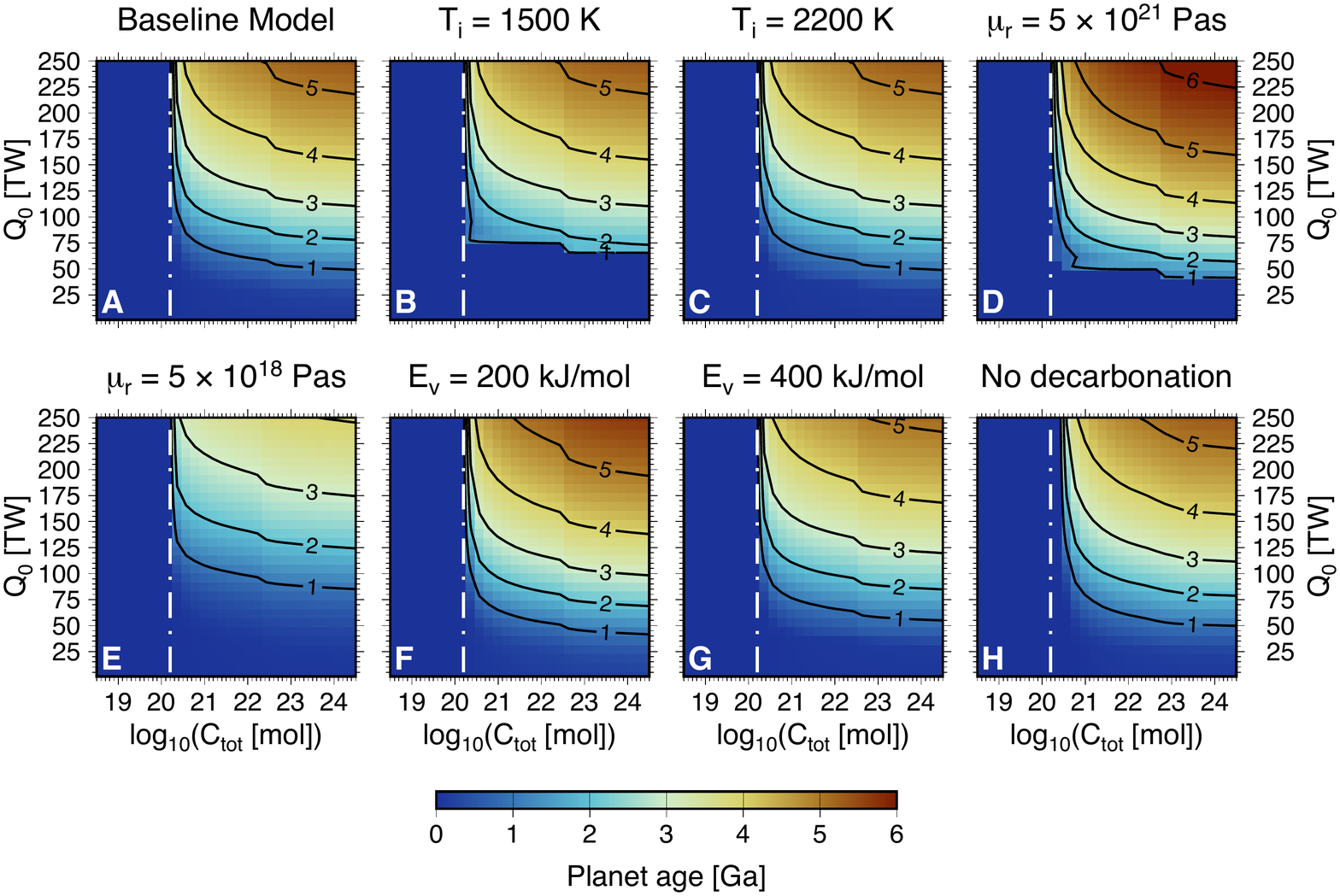}
\caption{\label {fig:rectime1_sens}  Planet age after which recovery from a soft snowball state becomes impossible, for planets with an initially hot climate and (A) baseline model parameters, (B) $T_i = 1500$ K, (C) $T_i = 2200$ K, (D) $\mu_r = 5 \times 10^{21}$ Pa s, (E) $\mu_r = 5 \times 10^{18}$ Pa s, (F) $E_v = 200$ kJ/mol, (G) $E_v = 400$ kJ/mol, (H) no crustal decarbonation. The white dot-dashed line marks the limit of $C_{tot} = 1.6 \times 10^{20}$ mol, below which recovery from a soft snowball state is never possible, due to there being insufficient CO$_2$ available in the system to melt an ice layer.  }
\end{figure} 

Similar effects of the key model parameters, discussed above, are seen when looking at planets' ability to recover from soft snowball states (Figure \ref{fig:rectime1_sens}). With a lower initial mantle temperature, planets with initial radiogenic heat production rates lower than {$\approx 75$} TW would never be able to recover from a soft snowball state, due to a lack of volcanism; at higher $Q_0$ the age where snowball recovery becomes impossible is similar to the baseline model (Figure \ref{fig:rectime1_sens}B). With $T_i = 2200$ K, results are indistinguishable from the baseline model (Figure \ref{fig:rectime1_sens}C). Consistent with the results for habitable climate lifetime, a higher reference viscosity extends the age range where snowball states can be recovered from, while a lower reference viscosity shortens this range (Figure \ref{fig:rectime1_sens}D \& E), as a result of how mantle viscosity influences mantle cooling rate. A lower activation energy for viscosity increases the planet age where snowball recovery becomes impossible, while a higher activation energy has the opposite effect, again due to the way activation energy influences mantle cooling rate (Figure \ref{fig:rectime1_sens}F \& G). Finally, when crustal decarbonation is assumed to not occur, the planet age when recovery from a soft snowball state is no longer possible is similar to the baseline model, that includes decarbonation (Figure \ref{fig:rectime1_sens}H). 

\begin{figure}
\includegraphics[width=1\textwidth]{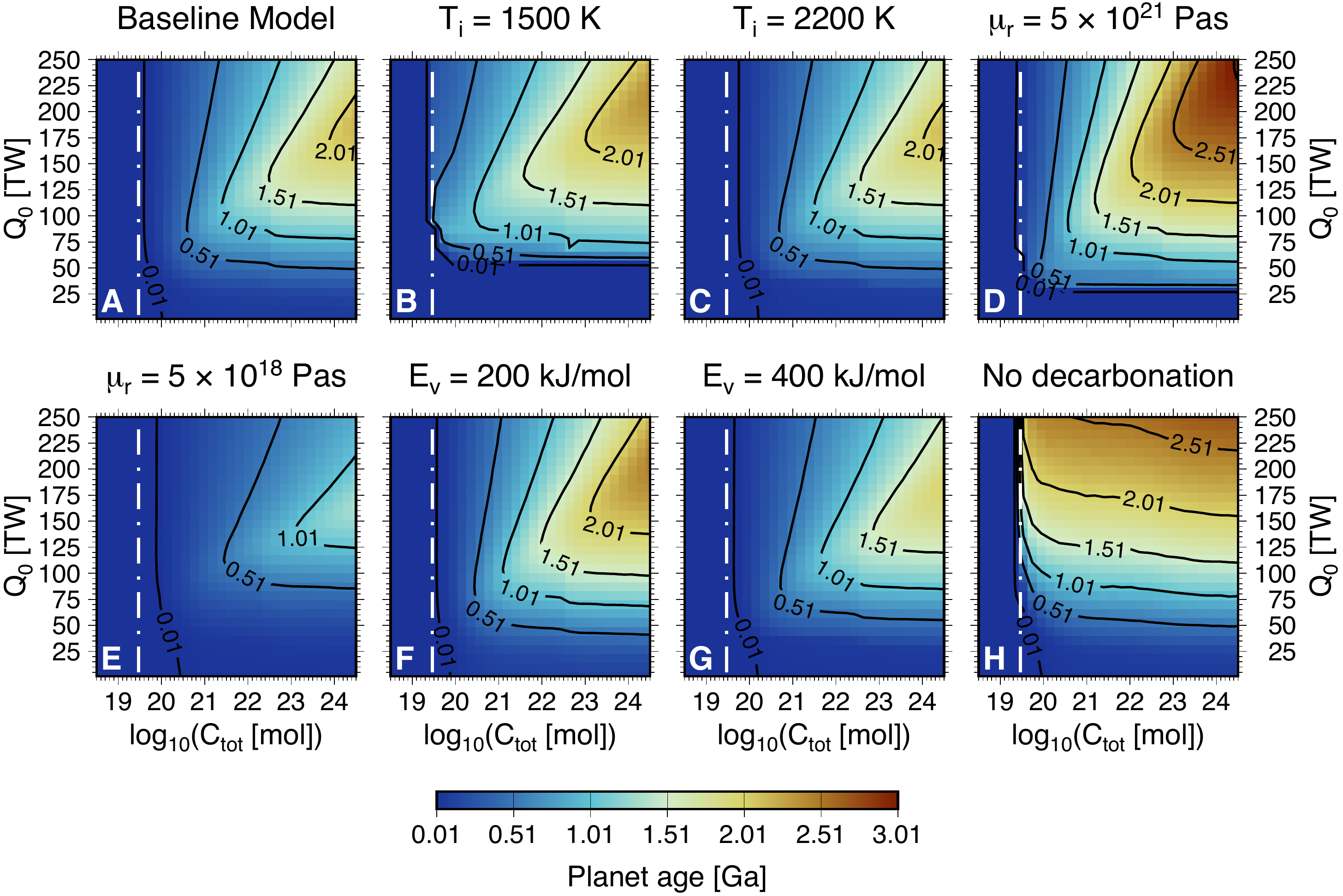}
\caption{\label {fig:rectime2_sens}  Planet age after which recovery from a hard snowball state becomes impossible, for planets with an initially cool climate and (A) baseline model parameters, (B) $T_i = 1500$ K, (C) $T_i = 2200$ K, (D) $\mu_r = 5 \times 10^{21}$ Pa s, (E) $\mu_r = 5 \times 10^{18}$ Pa s, (F) $E_v = 200$ kJ/mol, (G) $E_v = 400$ kJ/mol, (H) no crustal decarbonation. The white dot-dashed line marks the limit of $C_{tot} = 3 \times 10^{19}$ mol, below which recovery from a hard snowball state is never possible. }
\end{figure}

The planet age when recovery from a hard snowball state is no longer possible is shown in Figure \ref{fig:rectime2_sens}, in this case for an initially cold climate, as with an initially hot climate recovery from a hard snowball is nearly always impossible. The results are similar to those for soft snowball recovery, with the same effects of initial temperature and viscosity. One notable case is when decarbonation of the crust is assumed to not occur; here the age when hard snowball recovery becomes impossible shows a very similar pattern in $C_{tot}-Q_0$ space to the soft snowball recovery case, but with lower ages throughout the parameter space (Figure \ref{fig:rectime2_sens}H). The reason for this is that without decarbonation of the crust, the mantle CO$_2$ reservoir is constantly resupplied by foundering of carbonated crust, and thus not quickly depleted as in the other models.   

\begin{figure}
\includegraphics[width=1\textwidth]{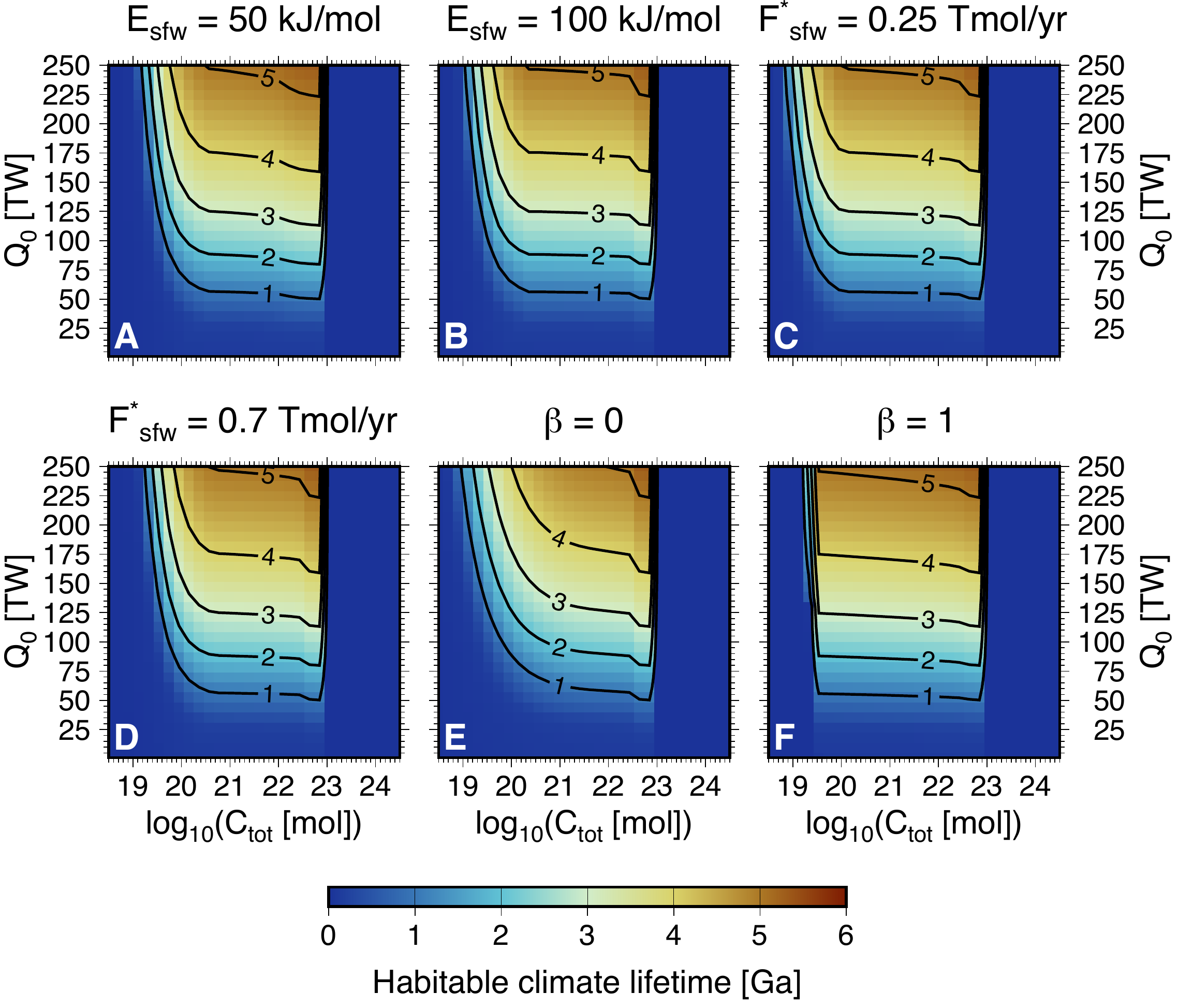}
\caption{\label {fig:habtime_sens2}  Lifetime of habitable climates for planets with an initially hot climate and (A) $E_{sfw} = 50$ kJ mol$^{-1}$, (B) $E_{sfw} = 100$ kJ mol$^{-1}$, (C) $F_{sfw}^* = 0.25$ Tmol yr$^{-1}$, (D) $F_{sfw}^* = 0.7$ Tmol yr$^{-1}$, (E) $\beta=0$, and (F) $\beta = 1$. }
\end{figure} 

Finally, the effect of uncertainties in the seafloor weathering function is explored (Figure \ref{fig:habtime_sens2}). Only the influence of the seafloor weathering parameters on the habitable climate lifetime is shown, as the maximum age when planets can recover from either a soft or hard snowball climate state is almost entirely unaffected by varying these parameters. Changing the seafloor weathering activation energy, $E_{sfw}$, does not significantly change the longevity of habitable climates, though with a larger $E_{sfw}$, habitable climates can last longer at low $C_{tot} \sim 10^{19}$ mol, as a result of more effective climate buffering. The reference seafloor weathering rate, $F_{sfw}^*$ has a similar effect; when $F_{sfw}^*$ is lower habitable climates can last longer at low values of $C_{tot}$ then when $F_{sfw}^*$ is higher, as the lower overall weathering rate means more CO$_2$ can build up in the atmosphere and keep the climate warm. The parameter with the biggest influence is the exponent describing the dependence of seafloor weathering on volcanic eruption rate, $\beta$. With $\beta = 1$, i.e. a strong dependence of seafloor weathering rate on volcanic eruption rate, planets with lower initial radiogenic heating rates and lower total CO$_2$ budgets ($C_{tot} < \sim 10^{20}$ mol) can maintain habitable climates for longer than when $\beta=1/2$ (the baseline value) or $\beta = 0$. The reason for this is that a strong dependence of seafloor weathering on eruption rate causes the seafloor weathering rate to decline as volcanism wanes; as a result, lower outgassing rates still allow enough CO$_2$ to remain in the atmosphere for a temperate climate to exist. 

\section{Discussion}

The results are generally consistent with \cite{Foley2018_stag}, which constrained the lifetime of potentially habitable climates on stagnant lid planets by calculating outgassing rates and estimating when supply limited weathering was expected to prevail. However, this study improves on the results of \cite{Foley2018_stag} by explicitly modeling atmospheric CO$_2$ abundance and climate evolution, to directly constrain when habitable climates can exist. As a result habitable climates are found to exist over a wider range of $C_{tot}$ than in \cite{Foley2018_stag}; $C_{tot} \sim 10^{19}-10^{23}$ mol, compared to $C_{tot} \sim 10^{20}-10^{22}$ mol. The reason for the difference is two fold. First, seafloor weathering results in generally lower CO$_2$ drawdown rates than continental weathering; the present day Earth seafloor weathering rate, $F_{sfw}^*$, determines the overall scale of the weathering rate, and $F_{sfw}^*$ is approximately a factor of 10 lower than the modern day continental weathering rate \citep{KT2017}. As a result, lower outgassing rates can still support a temperate climate, as the low CO$_2$ drawdown rate allows more CO$_2$ to remain in the atmosphere. A consequence of the slower rate of seafloor weathering is that this study finds outgassing rates need only be $\sim 10$ \% Earth's present day degassing rate to sustain temperate climates, consistent with the lower bound explored in \cite{Foley2018_stag} and with \cite{Haqq2016}. Second, while supply limited weathering begins at $C_{tot} \sim 10^{22}$ mol in this study, consistent with \cite{Foley2018_stag}, uninhabitabily hot climates do not develop until $C_{tot} \sim 10^{23}$ mol. That is, planets must be well within the supply limited weathering regime before sufficient CO$_2$ builds up in the atmosphere to result in extremely hot climates. Thus, the prospects for habitability of stagnant lid planets are even better than previously estimated. 

There are important caveats, however. As highlighted in \S \ref{sec:hysteresis}, at low CO$_2$ budgets, $C_{tot} < \sim 10^{20}$ {mol}, planets would be incapable of ever recovering from a snowball state, including just after their formation. Thus these planets would experience drastically different climate evolutions based on their initial conditions; if they start in a snowball state, this snowball will be permanent and the planet will never be habitable for surface life, while if they start with a temperate or hot climate, then a habitable climate can persist for $\approx 1$ Gyr or more, depending on the radiogenic heating rate. Planets with these low CO$_2$ budgets also would be unable to recover from a snowball climate state if it developed later in the planet's history as well. However, planets with larger CO$_2$ budgets, $C_{tot} > \sim 10^{20}$, mol can recover from initial snowball states. Such planets are found to be capable of recovering from soft snowball states, should one develop, throughout nearly their entire lifetime spent with significant outgassing and a temperate climate. However, hard snowball states are much more difficult to recover from, and planets developing such a climate will often be stuck in this state for the remainder of their lifetime. 

The initial distribution of carbon between mantle and atmosphere is thus important for planetary climate evolution. This distribution is highly uncertain, but most studies of magma ocean solidification indicate that this process depletes the mantle, and enriches the atmosphere, in CO$_2$ \citep[e.g.][]{Zahnle2007,Lindy2008,Hamano2013,Salvador2017}. Thus planets that have initially hot climates are probably more common, as high energy collisions during planet formation are expected to lead to magma oceans \citep{Lindy2012}. As a result, initial snowball states that become permanent are probably rare for the ``typical" stagnant lid planet, though ``typical" stagnant lid planets would be unable to recover from a hard snowball state, should such a state develop later during the planet's evolution. An important implication of the hysteresis in climate evolution is that some planets that would be predicted to be habitable and possess temperate climates, based on their CO$_2$ and heat budgets, may in fact lie in permanent snowball states. 
	
Another important caveat is that changes in solar luminosity are not considered in this study. By using a climate parameterization based on the present day Earth, a present day solar luminosity is implicitly assumed. In reality, stars are dimmer earlier in their history, and luminosity increases over time \citep[e.g.][]{Gough1981}. This change in luminosity could impact both the lifetime of habitable climates predicted by the thermal evolution models, and the ability of stagnant lid planets to recover from snowball climates. In particular, planets could be colder than our models predict early in their history, and warmer later on. However, seafloor weathering does include a strong negative feedback that acts to stabilize climate in the face of varying solar luminosity \citep{Brady1997,Coogan2013,Coogan2015,KT2018}. In fact, seafloor weathering can buffer climate against changes in luminosity more effectively than continental weathering, because seafloor weathering contains only a very weak direct dependence on atmospheric CO$_2$, through changes in ocean pH, and is instead much more sensitive to temperature; this results in stronger buffering with solar luminosity \citep{KT2018}. The main effect changes in solar luminosity would have on the habitable lifetimes calculated here is that at low luminosity, larger $p\mathrm{CO}_2$ is needed in order to keep the climate temperate. Thus when $C_{tot}$ is low enough that there is not enough carbon available to meet the required $	p\mathrm{CO}_2$ for a temperate climate, habitability will not be possible until luminosity increases. This threshold value of $C_{tot}$ is expected to be higher with a lower luminosity than the $C_{tot} \sim 10^{19}$ threshold we find in the models presented here. Likewise, if luminosity is higher, then lower values of $C_{tot}$ are expected to result in uninhabitability hot climates, than what is presented in this study. 

Solar luminosity also has an important influence on recovery from snowball climates. With a lower luminosity, more CO$_2$ in the atmosphere is required to melt the ice layer, while with a higher luminosity less CO$_2$ is required. Thus, recovery from both a hard and a soft snowball state will be harder when luminosity is lower, and easier when luminosity is higher. In particular, the value of $C_{tot}$ required to recover from either a hard or soft snowball climate will increase when luminosity is lower, and decrease when luminosity is higher. Fortunately, the effect of luminosity on snowball recovery can easily be captured with the models presented here, as it was found that the limit where a snowball state can never be recovered from is simply when the CO$_2$ budget is lower than what is needed to melt an ice layer. Thus, one could determine different limits on $C_{tot}$ needed to recover from snowball states at different solar luminosities, and the models in this study would indicate that planets with CO$_2$ budgets below this limit will never be able to recover from a snowball.  

As seen in \S \ref{sec:sensitivity}, the model results are largely consistent over a wide range of parameter values. Moreover, \cite{Foley2018_stag} performed even more extensive testing, in particular looking at the influence of incomplete degassing of both the crust and mantle, and found that these effects also do not significantly impact the results. I note that as the thermal evolution model and supply limited weathering formulation used in this study are similar to \cite{Foley2018_stag}, the same limitations of this model discussed in \cite{Foley2018_stag} apply here. There are interesting implications of how some of the key parameters influence the prospects for habitability of stagnant lid planets, however. In particular, the higher the mantle reference viscosity, or lower the activation energy, the longer volcanism lasts and the longer habitable climates persist. Thus planets with mantle mineralogies that lead to larger overall viscosities, or lower activation energies, compared to Earth would be better homes for life. Ultimately the models of stagnant lid planet habitability presented here could be used to constrain the likelihood of finding temperate climates on exoplanets as they are {discovered}. In particular if abundances of radionuclides and major rock forming elements in the star can be measured, and used to constrain the composition of the planet, direct estimates of the likelihood of habitability can be made, providing a valuable tool in the search for extra-terrestrial life.

\section{Conclusions}

Models coupling mantle thermal evolution, volcanism, outgassing, CO$_2$ cycling, and climate evolution for Earth-sized stagnant lid planets constrain the conditions that lead to long-lived temperate climates on such planets. Specifically, the lifetime of habitable climates is calculated as a function of the initial radiogenic heating budget of the mantle and total CO$_2$ budget of the mantle and surface reservoirs. Planets with CO$_2$ budgets of $C_{tot} \sim 10^{19}-10^{23}$ mol are found to {sustain} habitable climates for 1-5 Gyrs. When $C_{tot}$ is lower than this limit planets will remain frozen over for their entire lifetime, and above this limit planets will remain uninhabitably hot. The initial radiogenic heating rate is also important; heating rates of $\approx 50$ TW are needed for habitable climates to last for at least 1 Gyr, and increasing the heating rate increases the lifetime of volcanism, and hence temperate climates. Stagnant lid planets are assumed to be dominated by seafloor weathering in this study, and the strong climate buffering ability of seafloor weathering, along with its lower overall weathering rate as compared to continental weathering on Earth, allow a wide range of CO$_2$ budgets to support long-lived habitable climates. 

The ability of stagnant lid planets to recover from potential snowball episodes is considered. If a hard snowball, where there is no exchange between atmosphere and ocean, develops, recovery is very difficult as most CO$_2$ outgassing occurs via metamorphic decarbonation of the crust, below the ice layer. With a soft snowball, where there is atmosphere-ocean exchange, recovery is possible under most conditions. However, both snowball scenarios place limits on the CO$_2$ budget a planet must possess in order to ever recover from a snowball state. Planets with very small CO$_2$ budgets, lower than $\sim 10^{19}-10^{20}$ mol, would be unable to recover from a snowball climate, even if this occurs at the start of the planet's evolution. Thus such planets could exhibit hysteresis in their evolution, where planets that start off with a snowball climate, or develop one later, will be stuck in these states permanently, while planets that start with a warm or hot climate will be able to sustain habitable climates for 1 Gyr or more. Hysteresis means that planets one might otherwise expect to be habitable, based on their radiogenic heat and CO$_2$ budgets, may instead be found in a permanent snowball state. However, it is not clear that planets are likely to become stuck in these snowball states, as magma ocean solidification tends to release mantle CO$_2$ to the atmosphere, and produce an initially warm climate. Overall, the results of this study can be used to estimate the planetary factors that allow for long-lived habitability, aiding in target selection for future missions searching for biosignatures on exoplanets.   

{
\acknowledgments

I thank an anonymous reviewer for comments that helped to significantly improve the manuscript. }

{
\appendix
\section{Crustal carbon reservoir treatment}
\label{sec:appendix}
The models presented in the main text treat the crustal carbon reservoir as a single reservoir with instantaneous mixing. However, in reality there is no mixing in the crust, as it is part of the rigid lid, and crust decarbonating had its CO$_2$ concentration set as that parcel of crust formed at the surface. Such a simplification greatly helps with formulating the model, and here I show that treating the crustal carbon reservoir as a single, well mixed reservoir, does not significantly affect the results presented in the main text. The metamorphic outgassing flux calculated as in the main text is nearly identical, throughout most of a stagnant lid planet's thermal evolution, to the metamorphic outgassing flux that results from assuming that the CO$_2$ concentration of decarbonating crust is set when that parcel of crust forms. 

The concentration of CO$_2$ in crust formed at time $t$ is $C_{crust} = F_{weather}(t)/f_m(t)$, while the velocity of downgoing crust is $w_{crust}(t) = f_m(t)/A_{surf}$. The age of crust that is decarbonating at any time $t$ can be determined by finding the crust age, $t_{crust}$, that satisfies
\begin{equation}
\delta_{carb} = \int_{t_{crust}}^t w_{crust} dt .
\end{equation}
Thus the metamorphic outgassing rate where crustal CO$_2$ concentration varies with depth, is given by 
\begin{equation}
F_{meta}(t)' = \frac{F_{weather}(t_{crust})f_m(t)}{2f_m(t_{crust})}(\tanh{((\delta_c(t)-\delta_{carb}(t))20)}+1) .
\end{equation}
The metamorphic outgassing rate as formulated in the main text, $F_{meta}$, and $F_{meta}'$ are compared in Figure \ref{fig:fmeta_comp} for a few representative models, while the decarbonation depth and velocity of downgoing crust, $w_{crust}$, is shown in Figure \ref{fig:crust_vel} for the same models. The two metamorphic outgassing rates are nearly identical for most of the period over which volcanism is active. There are differences in outgassing rate very early in planet evolution, in particular for the cases where the mantle viscosity is high or initial mantle temperature is low; in these cases the mantle must heat up before volcanism can begin, and hence the start of CO$_2$ outgassing is delayed. In either formulation, metamorphic outgassing begins at the same time, as even with the formulation in the main text where instantaneous mixing in the crustal CO$_2$ reservoir is assumed, metamorphic outgassing can not take place until the crust first extends deep enough to reach the decarbonation depth. Also note that the crust is always at least somewhat carbonated, as all crust that forms experiences weathering as it is deposited at the surface. 

Very early in model evolution there can be significant differences between the two outgassing formulations as the weathering rate is either initially rapid (for a hot start initial condition) or sluggish (for a cold start). In particular, with $T_i = 1500$ K or $\mu_r = 5 \times 10^{21}$ Pa s, there is a rapid draw down of CO$_2$ that causes a cold climate, which then warms as metamorphic outgassing ramps up. During this cold climate phase, the CO$_2$ concentration in the crust is reduced, and hence $F_{meta}'$ shows a short-lived dip. However, at later times for all models, the majority of the planet's CO$_2$ has been outgassed from the mantle and is circulating through the crust. Over long time scales, the weathering rate balances the outgassing rate, meaning that the concentration of CO$_2$ in newly created crust is $F_{meta}'/f_m$. With $F_{meta}(t)' = C_{crust}(t_{crust})f_m(t)$, the concentration of CO$_2$ in newly created crust is the same as the concentration of CO$_2$ in the crust at the decarbonation depth. Thus once metamorphic outgassing becomes the dominant CO$_2$ source to the atmosphere, the concentration of CO$_2$ in the crust is constant, and the outgassing rate simply declines as the melt production rate declines. While the simple formulation of metamorphic outgassing used in the main text might miss some interesting changes in outgassing, and hence climate, early in a planet's history, it is a good approximation for the metamorphic outgassing rate for most a planet's history. The overall results of this study, which focus on the longevity of habitable climates on stagnant lid planets, are therefore unlikely to be significantly affected by the assumption of an instantaneously mixed crustal carbon reservoir.  

\begin{figure}
\includegraphics[width=0.5\textwidth]{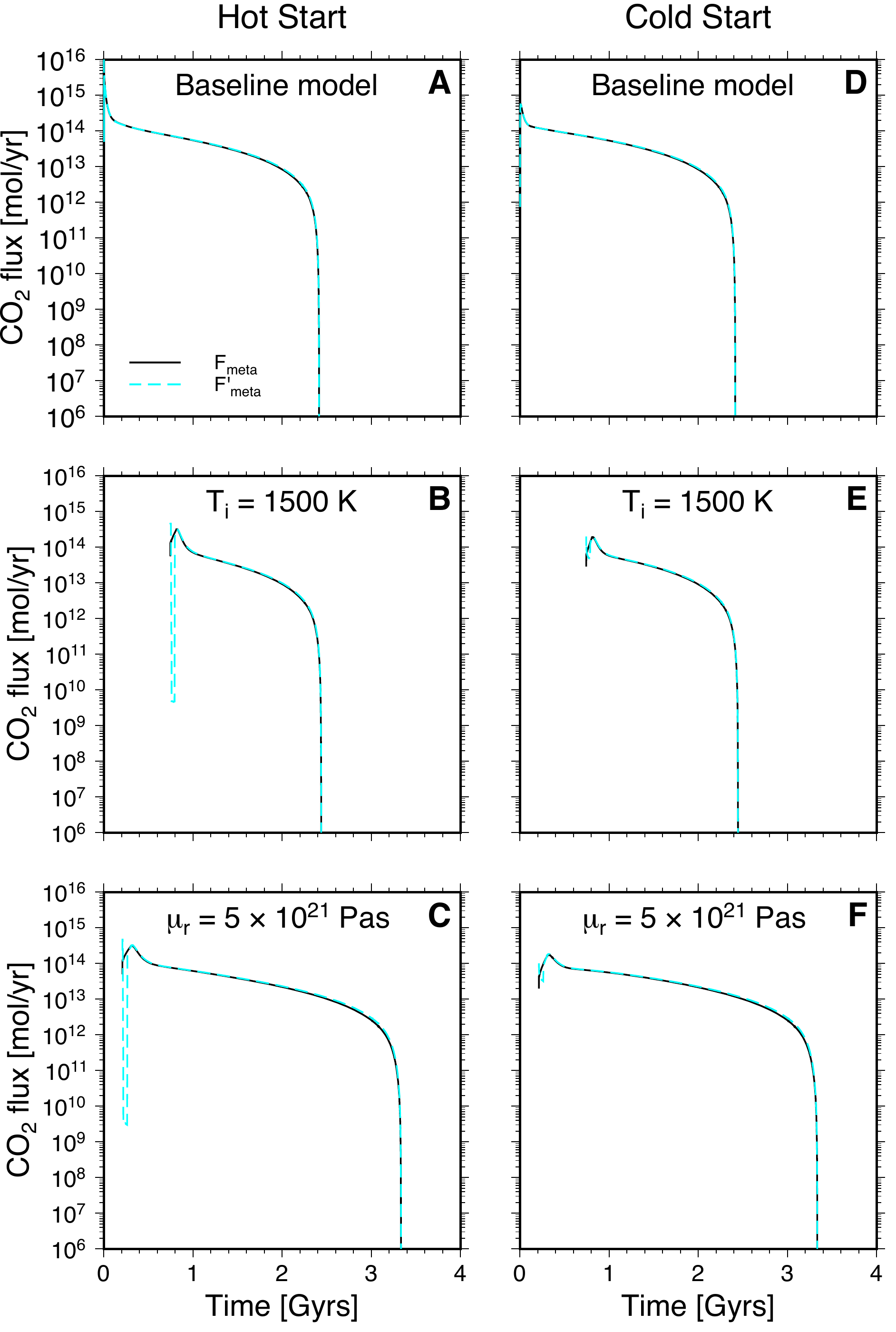}
\caption{\label {fig:fmeta_comp}  Comparison between $F_{meta}$ and $F_{meta}'$ for models with $C_{tot} = 10^{22}$ mol, $Q_0 = 100$ TW, and baseline parameter values (A \& D), with $T_i = 1500$ K (B \& E), and with $\mu_r = 5 \times 10^{21}$ Pa s (C \& F). A model with an initially hot climate and one with an initially cold climate are run for each set of parameters.  }
\end{figure} 

\begin{figure}
\includegraphics[width=0.5\textwidth]{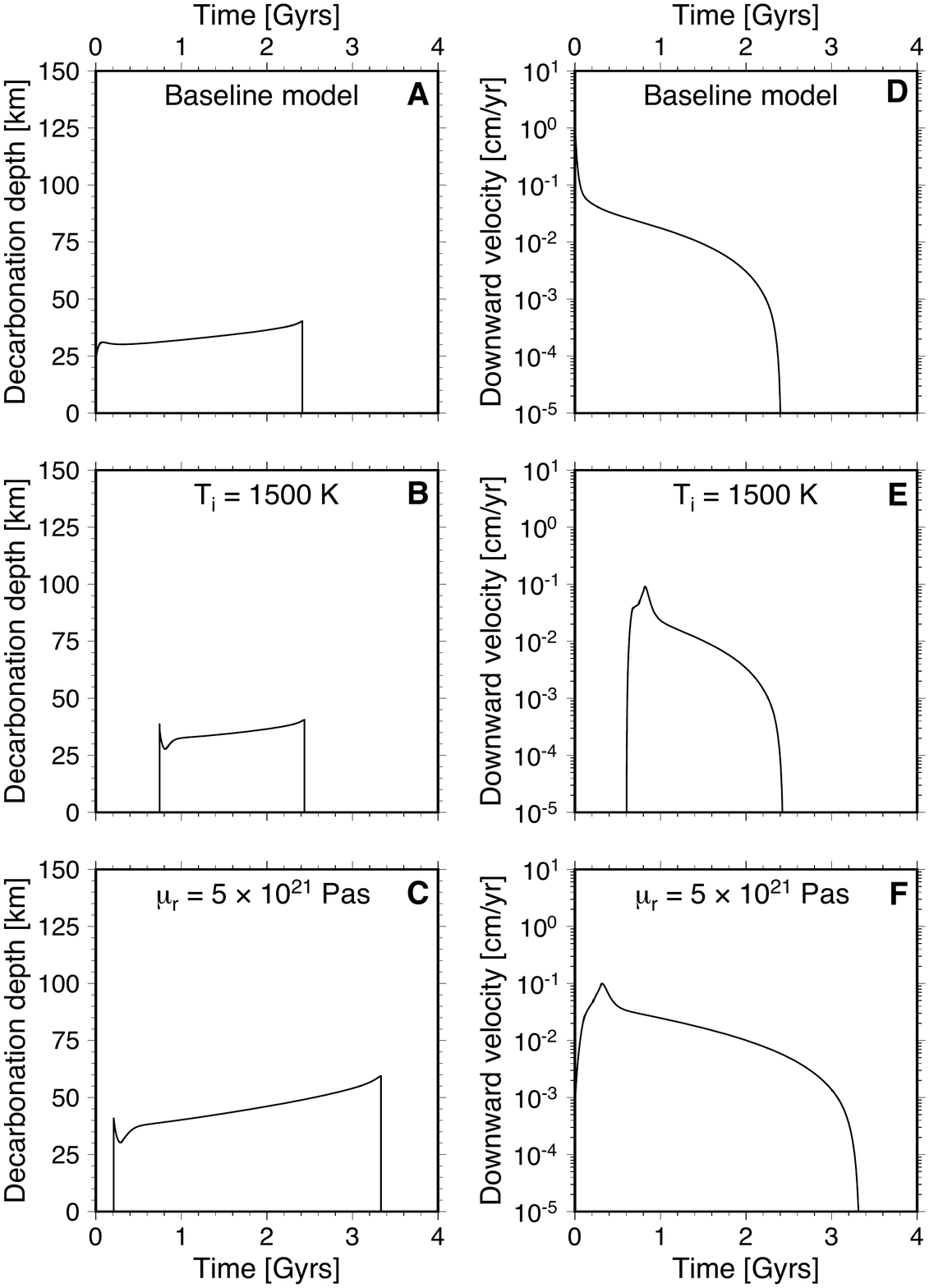}
\caption{\label {fig:crust_vel}  Metamorphic decarbonation depth (shown only for the time period when metamorphic decarbonation is active) and downward crustal velocity for models with $C_{tot} = 10^{22}$ mol, $Q_0 = 100$ TW, and baseline parameter values (A \& D), with $T_i = 1500$ K (B \& E), and with $\mu_r = 5 \times 10^{21}$ Pa s (C \& F). All models shown assume an initially hot climate.  }
\end{figure} 

}

\clearpage

\def\els{/Users/bfoley/Documents/tex/elsart}
\def\agu{/Users/bfoley/Documents/tex/agu}
\def\bib{/Users/bfoley/Documents/tex/bib}

\bibliographystyle{model5-names}
\bibliography{GeneralGeophys+Misc,MantleConvection,TreatiseXtra,seismology,TwoPhase-Damage,planetaryscience,geochem,bibtex} 

\clearpage

\end{document}